\documentclass[11pt,a4paper]{article}
\usepackage{jheppub,bm,color}
\usepackage{xcolor}
\usepackage{amsmath,amssymb,bm}
\usepackage{lineno}

\usepackage[normalem]{ulem}  

\def\lcb{l_\text{Cb}}
\def\lcbdot{\dot{l}_\text{Cb}}
\def\be{\begin{equation}}
\def\ee{\end{equation}}
\def\bes{\begin{subequations}}
\def\ees{\end{subequations}}

\def\no{\nonumber}
\def\a{\alpha}
\def\b{\beta}

\def\g{\gamma}

\def\p{\psi}

\def\s{\sigma}

\def\vp{{\bm p}}

\def\pd{\partial}

\def\le{\left}
\def\ri{\right}

\def\<{\langle}
\def\>{\rangle}
\def\+{\dagger}

\def\[{\left[}
\def\]{\right]}

\def\C{{\mathcal C}}

\def\I{{\mathcal I}}
\def\p{{\bf p}}

\def \pT{p_{\perp}}
\def \pz{p_{z}}

\def\tA{A}

\def\ta{a}

\def\sE{{\cal E}}

\def\tq{\widetilde{q}}
\def \tqB {\widetilde{q}_B}

\def\ad{\text{ad}}

\def\als{\alpha_{S}}
\def\betas{\beta_{S}}
\def\gas{\gamma_{S}}

\def\As{A_{S}}
\def\Bs{B_{S}}
\def\Cs{C_{S}}
\def\qs{q_{S}}

\begin{document}

\preprint{CERN-TH-2022-026, MIT-CTP/5411}

\title{
Scaling and adiabaticity in a rapidly expanding gluon plasma
}

\author[a]{Jasmine Brewer}
\author[b]{Bruno Scheihing-Hitschfeld}
\author[c]{and Yi Yin}

\affiliation[a]{Theoretical Physics Department, CERN, CH-1211 Gen\`eve 23, Switzerland}

\affiliation[b]{
Center for Theoretical Physics, Massachusetts Institute of Technology, Cambridge, Massachusetts 02139, USA
}

\affiliation[c]{
Quark Matter Research Center, Institute of Modern Physics, Chinese Academy of Sciences, Lanzhou, Gansu, 073000, China
}

\abstract{
In this work we aim to gain qualitative insight on the far-from-equilibrium behavior of the gluon plasma produced in the early stages of a heavy-ion collision. It was recently discovered~\cite{Mazeliauskas:2018yef} that the distribution functions of quarks and gluons in QCD effective kinetic theory (EKT) exhibit self-similar ``scaling'' evolution with time-dependent scaling exponents long before those exponents reach their pre-hydrodynamic fixed-point values. In this work we shed light on the origin of this time-dependent scaling phenomenon in the small-angle approximation to the Boltzmann equation. We first solve the Boltzmann equation numerically and find that time-dependent scaling is a feature of this kinetic theory, and that it captures key qualitative features of the scaling of hard gluons in QCD EKT. We then proceed to study scaling analytically and semi-analytically in this equation. We find that an appropriate momentum rescaling allows the scaling distribution to be identified as the instantaneous ground state of the operator describing the evolution of the distribution function, and the approach to the scaling function is described by the decay of the excited states. That is to say, there is a frame in which the system evolves adiabatically. Furthermore, from the conditions for adiabaticity we can derive  evolution equations for the time-dependent scaling exponents. In addition to the known free-streaming and BMSS fixed points, we identify a new ``dilute'' fixed point when the number density becomes small before hydrodynamization. Corrections to the fixed point exponents in the small-angle approximation agree quantitatively with those found previously in QCD EKT and arise from the evolution of the ratio between hard and soft scales. 
}

\emailAdd{jasmine.brewer@cern.ch}
\emailAdd{bscheihi@mit.edu}
\emailAdd{yiyin@impcas.ac.cn}

\maketitle


\section{Introduction}

Exploring the many-body physics of Quantum Chromodynamics (QCD) is one of the key frontiers in high-energy nuclear physics. 
The last decade saw significant progress in understanding thermal and transport properties of the quark-gluon plasma (QGP) produced in heavy ion collisions (see e.g.~\cite{akiba2015hot,Romatschke:2017ejr,Busza_2018,Schenke:2021mxx} for reviews). Recently the far-from-equilibrium behavior of the QGP and the subsequent process of thermalization have attracted significant interest.
On the theory front, the non-equilibrium dynamics of QCD and QCD-like theories have been investigated using kinetic theory, classical field simulations,
and gauge/gravity duality (see~\cite{Schlichting:2019abc,Berges:2020fwq} for recent reviews).
There is a growing appreciation that the off-equilibrium dynamics in the early stages of heavy ion collisions may be crucial for understanding the observed collectivity~\cite{Kurkela:2018wud,Schenke:2019pmk,Giacalone:2020byk,Kurkela:2021ctp}, especially in smaller collision systems. 
QCD effective kinetic theory (EKT)~\cite{Arnold:2002zm} has become a particularly important tool for understanding far-from-equilibrium dynamics and thermalization in QCD (see e.g.~\cite{Kurkela:2015qoa,Kurkela:2018xxd,Almaalol:2020rnu,Du:2020zqg}).

Studies of far-from-equilibrium QCD have revealed a surprising self-similar ``scaling'' behavior of the quark and gluon distribution functions.
A distribution function is said to exhibit scaling behavior if the shape of the (rescaled) distribution function remains stationary when expressed in terms of a rescaled momentum variable.
This self-similar evolution has been observed in classical field simulations~\cite{Berges:2013eia,Berges:2013fga}, for small-angle scatterings in kinetic theory~\cite{Tanji:2017suk},
and later in simulations of QCD EKT~\cite{Mazeliauskas:2018yef}. 
In addition to showing that the distribution function reaches the self-similar scaling form, the study of Ref.~\cite{Mazeliauskas:2018yef} further demonstrated that
the distribution function can take the scaling form with time-dependent scaling exponents much before the scaling exponents attain their fixed-point (time-independent) values.\footnote{
In Ref.~\cite{Mazeliauskas:2018yef}, this time-dependent scaling is called ``prescaling".} 
This interesting and important finding suggests that the early time evolution of quark-gluon matter created in heavy-ion collisions might be simply characterized by a scaling function together with the evolution of a handful of scaling exponents. Time-dependent scaling in Bose gases has also been studied in Ref.~\cite{Schmied_2019}. However, it remains unclear what causes the emergence of time-dependent scaling and how general the resulting exponents are.

In this work we concentrate on the early non-equilibrium stage of a heavy-ion collision and aim at gaining qualitative lessons on the emergence of time-dependent scaling and the evolution of the scaling exponents.
For this purpose, 
we shall consider a Bjorken-expanding gluon plasma and study the kinetic Boltzmann equation with a highly occupied initial condition. 
We will employ the small angle approximation to the Boltzmann equation, which then takes the form of a Fokker-Planck (FP) equation. We will refer to this equation as ``FP equation'' throughout. 
This equation has been studied previously in Ref.~\cite{Tanji:2017suk}, where they showed that it featured solutions with time-independent scaling behavior. 
Our results further demonstrate that this FP equation exhibits \textit{time-dependent} scaling for hard gluons, and that its solutions capture key qualitative features of the scaling seen from QCD EKT results reported in Ref.~\cite{Mazeliauskas:2018yef}. 
This supports our view that the relatively simple FP equation can be utilized as a qualitatively accurate effective description of time-dependent scaling behavior of hard gluons.

\begin{figure}[t]
\centering
\includegraphics[width=0.9\textwidth]{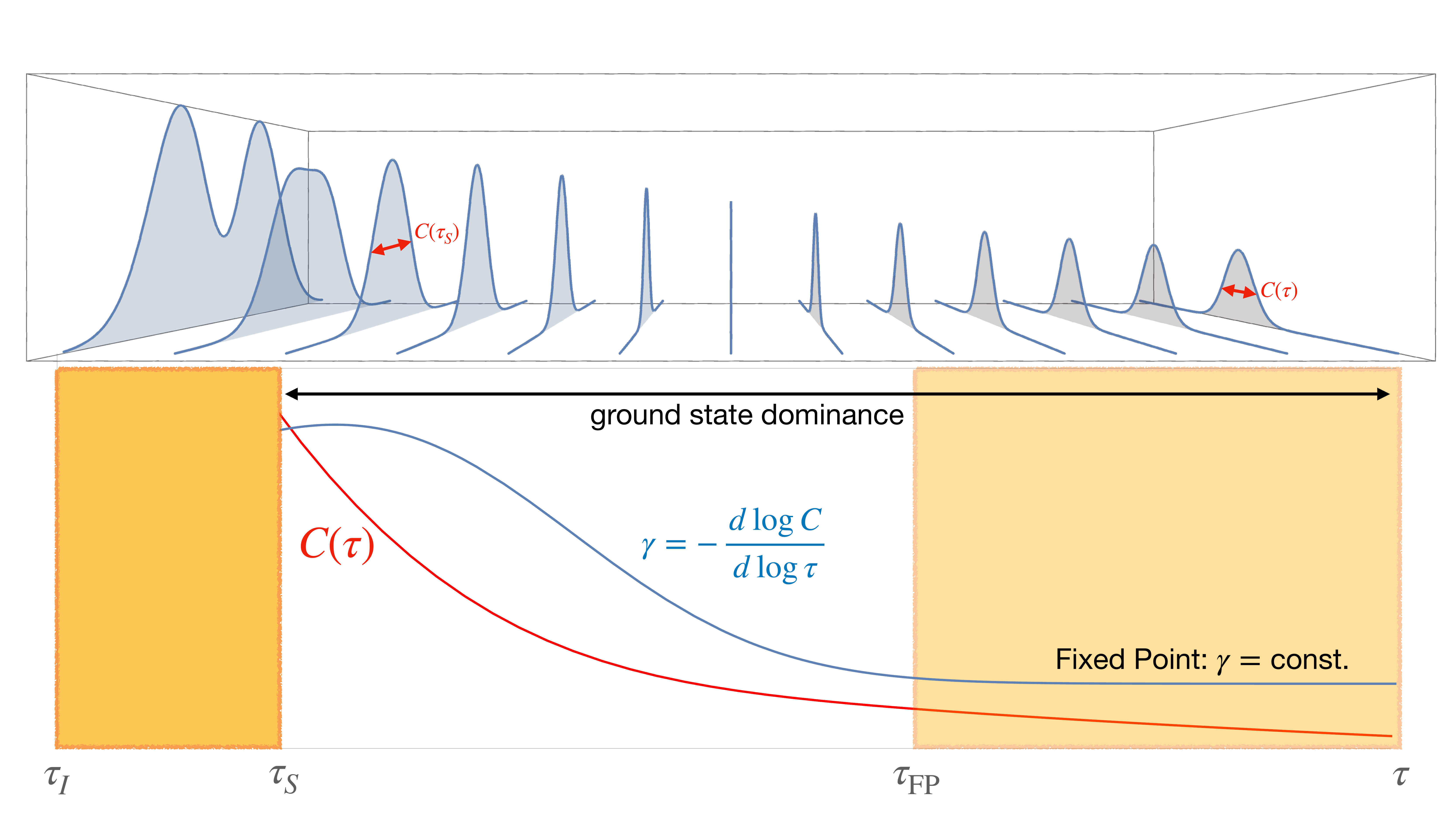}
  \caption{
  \label{fig:Evo-Ill}
We illustrate time-dependent scaling behavior and its connection to adiabaticity. On the top, we show a typical evolution of a distribution function in the present work.
Below, we show the temporal evolution of the characteristic scale $C$ and its associated scaling exponent $\gamma$ in red and blue solid curves, respectively. Though the evolution of the distribution function begins at $\tau_{I}$, the scale and exponents are only well-defined after the time $\tau_S$ when the distribution function reaches its self-similar scaling form.
Although the scaling exponent $\gamma$ will eventually approach its fixed point value at 
$\tau_{\rm FP}$, 
the distribution function may take the scaling form at $\tau_{S}<\tau_{\rm FP}$.
Within the present set-up, we find the emergence of scaling behavior around $\tau_{S}$ is associated with the decay of excited modes, as will be explained throughout the main text. 
The ground state mode can then be associated with the scaling form of the distribution, giving the dominant contribution to the state of the system during the scaling stage, and hence the distribution function's self-similar evolution becomes equivalent to adiabatic evolution. We note that, in this scenario, all of this happens well before the system becomes hydrodynamic: $\tau_{\rm FP} \ll \tau_{\rm Hydro}$.} 
\end{figure}

One of the novel results in this paper is that 
for the case of the FP equation, 
we show explicitly and analytically the equivalence between scaling in the evolution of the distribution function and the adiabatic evolution of the distribution function by extending the adiabatic scenario for rapidly expanding gluon plasmas first proposed in Ref.~\cite{Brewer:2019oha}. 
In our adiabatic picture, the scaling function can be identified as the instantaneous ground state of a non-Hermitian operator describing the evolution of a rescaled distribution function. 
In this framework, the emergence of self-similarity is due to the decay of instantaneous excited states. Excited states naturally decay over time because the non-Hermitian nature of the time evolution operator considered herein implies that their time evolution factors decay exponentially over time, like the evolution of states in quantum mechanics under Euclidean time evolution. 
The time scale over which this decay happens is determined by the inverse of the energy gap $\Delta E$ between the ground and lowest excited state. If this energy gap is larger than the rate $\Gamma_{0 \to e}$ at which transitions induced by the time-dependence of the time evolution operator move the system away from the ground state, then one says that the evolution is adiabatic, and furthermore, it is a good approximation to describe the evolution of the whole system by that of its ground state.
This naturally explains why a wide range of initial distributions would approach the scaling function, and showcases the generality of time-dependent scaling (see Fig.~\ref{fig:Evo-Ill} for an illustration).
In general, whenever such a description can be set up, this provides a simple and straightforward way to understand the emergence of pre-hydrodynamic attractors.

To describe the evolution of the scaling exponents,
we derive a set of closed-form equations by imposing the adiabaticity condition $ \Gamma_{0 \to e} / \Delta E \ll 1$ for the rescaled distribution function, which in the case of the FP equation we study can be made exact by demanding $\Gamma_{0 \to e} = 0$, thus ensuring that the state cannot transition away from the ground state.   
We verify numerically that those equations not only give a reasonable description for scaling exponents extracted from the FP equation, but also for those from QCD EKT~\cite{Mazeliauskas:2018yef}.
From those equations, 
we obtain non-universal corrections to the scaling exponents near the fixed point, analogous to anomalous dimension corrections in quantum field theory.

This work is organized as follows: 
we review the pertinent ingredients of time-dependent scaling behavior and specify the FP equation we solve throughout this work in sec.~\ref{sec:scaling}. 
Then, in sec.~\ref{sec:scaling-num}, we demonstrate the scaling behavior of the FP equation numerically, and in sec.~\ref{sec:analytic} present analytical results for the scaling solution for a simplified case in order to gain some intuition. Next, in sec.~\ref{sec:Adiabatic} we establish the connection between adiabaticity and scaling, demonstrating our claim that describing the distribution function in terms of an adiabatic evolution of the system makes manifest the underlying phenomena that lead to attractor behavior. In sec.~\ref{sec:evo}, we formulate and study the evolution equations for the scaling exponents that define the frame in which adiabaticity is optimized, and compare with available numerical results to test our formalism.
We give our concluding remarks in sec.~\ref{sec:conclusion}.

\section{Set-up} 
\label{sec:scaling}

In this work, 
we consider the early-time, far-from-equilibrium evolution of gluonic matter created in a heavy-ion collision undergoing Bjorken expansion.
We shall assume that the initial gluon distribution is given by the saturation scenario (see Ref.~\cite{Gelis:2010nm} for a review), i.e., the typical gluon momentum is the saturation scale $Q_{s}$ and the occupation number of hard gluons is much larger than $1$.
The gluon distribution will subsequently evolve because of the longitudinal expansion and interactions among gluons. 
Within the above picture,
we will investigate how a self-similar evolution of the gluon distribution function $f(\pz,\pT;\tau)$ (which depends on transverse and longitudinal momentum $p_{\perp}, p_{z}$ and Bjorken time $\tau$) can emerge.

In this section, we will establish the concepts we will need in our subsequent analysis. Specifically, we review pertinent ingredients of time-dependent scaling in subsection~\ref{sec:scaling-review} and specify the collision integral we use in subsection~\ref{sec:kin-setup}.

\subsection{Time-dependent scaling
\label{sec:scaling-review}
}

Let us begin by writing an arbitrary distribution function $f(\pz,\pT;\tau)$ as 
\begin{align}
\label{f-w}
    f(\pT,\pz; \tau)= \tA(\tau)\, w(\zeta,\xi; \tau)\, , 
\end{align}
where we have introduced the rescaled variables
\begin{align}
\label{zeta-xi-def}
  \zeta \equiv \frac{p_{\perp}}{B(\tau)}\, , 
  \qquad
  \xi \equiv \frac{p_{z}}{C(\tau)}\, .
\end{align}
Given that the function $w$ is time-dependent at this point, there is no loss of generality as any function $f$ can be written in this way. For simplicity in the notation, we shall henceforth keep the time-dependence of $A,B,C$ implicit.

The choice of $A,B,C$ can be viewed as a choice of frame. For a given distribution function $f$, there is a family of frames resulting in a family of rescaled distribution functions $w$. Though $A,B,C$ at this point are arbitrary, an appropriate frame choice may illuminate the underlying physics.\footnote{See also Refs.~\cite{2001nlin.....11055A,kevrekidis2017revisiting} for examples in the study of self-similar solutions for partial differential equations.} For convenience, 
we shall take $A,B,C$ to be order of the characteristic occupancy number, transverse, and longitudinal momentum respectively, so that $w$ is order one for $\zeta,\xi\sim 1$. Furthermore, if one is able to find a frame such that these properties are preserved under time evolution, then a great reduction in complexity is achieved because the characteristic scales of the problem are immediately apparent. Finding such a frame is one of the main tasks that we will undertake throughout the rest of this work.

The evolution of $A,B,C$ can be characterized by their percentage rate of change,
 \begin{align}
   \label{exp-def}
   \dot{A} \equiv \frac{\tau \pd_{\tau}A}{A} = \a(\tau)\, , 
   \qquad
   \dot{B}= -\beta(\tau)\, , 
   \qquad
   \dot{C}= -\g(\tau)\, . 
 \end{align}
Throughout this work, we will use the ``dot'' to denote the logarithmic derivative with respect to $\log\tau$, e.g., $\dot{X} \equiv \partial_{\log \tau} \log X$, and keep the time-dependence of $\a,\b,\g$ implicit unless otherwise specified.

To gain intuition for these changes of frames, and what to expect for the values of the scaling exponents throughout the system's evolution, we note that for a plasma undergoing rapid longitudinal expansion, the characteristic longitudinal momentum $C$ should drop as $1/\tau$ in the free-streaming limit, corresponding to $\g=1$. 
Once interactions become relevant, one expects that the momentum exchange among gluons would slow the decay of $C$, so we expect $0< \g < 1$. 
On the other hand, the change of the characteristic transverse momentum $B$ is solely due to interactions and hence is slower than that of $C$. 
This implies that generically during the early stages of the evolution we will have
\begin{align}
\label{BC-ratio}
    r\equiv\frac{C}{B}\ll 1\, , 
    \qquad
    \frac{|\beta|}{|\g|}\ll 1\, ,
\end{align}
(see also Ref.~\cite{Baier:2000sb}). 
When the collision integral is dominated by momentum diffusion, the width of the transverse momentum distribution broadens and we expect that $\beta \leq 0$.

A distribution function is said to exhibit \textit{scaling} if there exists a special (time-dependent) frame $\As,\Bs,\Cs$ in which $w$ becomes time-independent, i.e.,
\begin{align}
w(\zeta,\xi;\tau)=w_{S}(\zeta,\xi) \, ,
\end{align}
and the distribution function takes the scaling form
\begin{align}
   \label{prescaling-1}
  f(p_\perp,p_z;\tau) = \As(\tau) \, w_{S}\left( \frac{p_\perp}{\Bs(\tau)},\frac{p_z}{\Cs(\tau)} \right)\, .
  \end{align}
The distribution function $f$ generally changes rapidly in a fast-expanding gluon plasma. Scaling is the special property that this time-dependence can be absorbed into that of $\As,\Bs,\Cs$ so that the shape of the gluon distribution in rescaled coordinates may evolve slowly or become stationary (as in \eqref{prescaling-1}).

Fixed points of the evolution are characterized by the special case that $\als,\betas,\gas$ in eq.~\eqref{exp-def} are time-independent, and therefore
\begin{align}
\label{ABC-scale}
    \As\sim \tau^{\als}\, , 
    \qquad
    \Bs\sim \tau^{-\betas}\, , 
    \qquad
    \Cs\sim \tau^{-\gas}\, . 
\end{align}
Because of eq.~\eqref{ABC-scale}, $\als,\betas, \gas$ are commonly referred to as the \textit{scaling exponents}. 
Different values of $\als,\betas, \gas$ specify different fixed points. 
For example, in the bottom-up thermalization scenario, 
the gluon plasma transits from the free-streaming fixed point $(\als,\betas,\gas)=(0,0,1)$ to the Baier-Mueller-Schiff-Son (BMSS) fixed point $(\als,\betas,\gas)=(-2/3,0,1/3)$ first found in~\cite{Baier:2000sb}.

It is conceivable that a distribution function could take the scaling form before it evolves to the fixed point. 
In this case, $w$ approaches the time-dependent scaling function $w_{S}$ while the scaling exponents $\als,\betas,\gas$ still change in time. 
Ref.~\cite{Mazeliauskas:2018yef} demonstrated that the gluon and quark distribution functions exhibit this time-dependent scaling (also called ``prescaling'') in numerical simulations of QCD effective kinetic theory (EKT). This observation suggests a surprising simplification in the far-from-equilibrium evolution of the distribution function. The goal of the present work is to gain qualitative insight into this behavior.\footnote{
The time-dependent scaling of a distribution function bears a certain similarity to the crossover phenomenon of a critical Ising system.
In this case, the critical exponents evolve as a function of temperature $T$ (and/or magnetic field) from the mean-field values to those of Wilson-Fisher fixed point as $T$ approaches the critical temperature~\cite{cha95}.
}

\subsection{Kinetic equation and the small angle approximation
\label{sec:kin-setup}
}

We will work in the weak coupling regime $ g^{2}_{s}f\ll 1$, with $g_{s}$ the coupling constant. 
In this regime the evolution of the distribution function can be described by 
the Boltzmann equation~\cite{Mueller:2002gd}
\begin{equation}
\label{kin}
  \pd_{\tau}\,f - \frac{p_z}{\tau} \partial_{p_{z}} f = - \C[f]\, ,
\end{equation}
where $\C$ is the collision integral.
As we mentioned in the Introduction, 
we shall employ the small angle scattering approximation to the collision integral, which, as the name suggests, assumes that gluons interact exclusively through small-angle elastic scatterings. 
Then, the collision integral is reduced to a Fokker-Planck-like diffusive kernel~\cite{Mueller:1999pi,Blaizot:2013lga}
\begin{equation}
	\label{eq:small-angle-kernel}
  \C_{{\rm FP}}[f] = -\lambda_0 \lcb [f] \left[ \I_a[f] \nabla_{\p}^2 f + \I_b[f] \nabla_{\p} \cdot \left( \frac{\p}{p} (1+f) f \right) \right]\, ,
\end{equation}
where $\lambda_0 = \frac{g^{4}_{s}}{4\pi} N_c^2$.
Throughout this work we refer to the Boltzmann equation~\eqref{kin} with the collision integral~\eqref{eq:small-angle-kernel} as the Fokker-Planck (FP) equation.  
The functionals $\I_a$, $\I_b$ are given by
\begin{align}
	\label{eq:IaIb}
  \I_a[f] = \int_{\vp} f (1+f), \, 
  \qquad
  \I_b[f] = \int_{\vp}\, \frac{2}{p} f\, , 
\end{align}
where here and throughout we use the shorthand notation $\int_{\vp} \equiv \int \frac{d^3 p}{(2\pi)^3}$. 
The integrand of $\I_{a}$ is proportional to the density of possible scatterers and hence will be enhanced by the Bose factor when $f>1$.
$\I_{b}$ is related to the Debye mass $m_{D}$, the typical momentum exchange per collision, by (see for example Ref.~\cite{Arnold:2008zu})
\begin{equation}
\label{mD}
  m_D^2 = 2 N_c g^2_{s} \I_b\, .
\end{equation}
The Coulomb logarithm $\lcb$ represents a (perturbatively divergent) integral over the small scattering angle~\cite{Mueller:1999pi} 
\begin{equation}
\label{lcb-def}
  \lcb [f] =\ln \left( \frac{p_{\rm UV}}{p_{\rm IR}} \right)\, ,
\end{equation}
where $p_{\rm UV}$ and $p_{\rm IR}$ are UV and IR cutoffs, respectively.
This IR divergence originates from the long range nature of the color force and is regularized by
the thermal medium-induced mass, so we take $p_{\rm IR}$ to be $m_D$. 
Since the distribution function has finite support in momentum space, 
we take $p_{\rm UV}$ to be the characteristic hard momentum of gluons above which the occupation number starts to decrease.
When the typical transverse momentum scale is much greater than the longitudinal momentum scale, which is the case when the medium is undergoing rapid longitudinal expansion (see eq.~\eqref{BC-ratio}),  
we use
\begin{align}
p_{{\rm} UV}=\sqrt{\langle p^{2}_{\perp}\rangle}\, ,
\end{align}
where the average over the distribution function is defined in a standard way 
\begin{align}
\label{average}
  \langle\ldots \rangle\equiv \frac{\int_{\vp}\, (\ldots)\,f}{\int_{\vp}\, f}\, . 
\end{align}
As a result, for our present purposes the expression for the Coulomb logarithm can be explicitly written as~\cite{Mueller:1999fp} 
\begin{equation}
\label{lcb-0}
  \lcb [f] = 
  \ln \left( \frac{\sqrt{\langle p^{2}_{\perp}\rangle}}{m_{D}} \right)\, .
\end{equation}
Since both $p_{\rm IR}$ and $p_{\rm UV}$ are functionals of the distribution function, they themselves are time-dependent, and therefore so is $\lcb$.
We will later demonstrate in sec.~\ref{sec:a-dim} that
the temporal dependence of $\lcb$ plays an interesting role in determining the precise behavior of the scaling exponents near the fixed points.

In the coming section, we will first establish the emergence of time-dependent scaling in the FP equation in the hard transverse momentum regime $\zeta \geq 1$ for all $\xi$. 
Gluons in this regime have typical longitudinal momentum much smaller than their typical transverse momentum, and therefore $r=C/B$ is small (see eq.~\eqref{BC-ratio}). 
This allows us to analyze the scaling behavior order by order in $r$. To the zeroth order in the small $r$ limit,
it is sufficient to consider only longitudinal momentum diffusion in the collision integral
\begin{equation}
\label{CIa-0}
    C[f] = -\lambda_0 \lcb[f] \I_a[f]\, \partial^2_{p_{z}} f \, .
\end{equation}
At finite $r$ we find that setting $\I_b=0$ in $C_{{\rm FP}}[f]$, i.e.
\begin{equation}
\label{CIa-1}
    C[f] = -\lambda_0 \lcb[f] \, \I_a[f] \, \nabla_{\bf p}^2 f \, ,
\end{equation}
accurately describes sufficiently hard gluons as long as $A>1$ (see Appendix~\ref{sec:Ib}).
This anticipation will be corroborated by the numerical calculations in sec.~\ref{sec:scaling-num}.
We therefore use eqs.~\eqref{CIa-0} and \eqref{CIa-1} for the analytic part of our study of self-similarity and the  scaling behavior of the distribution function.

Before closing this section, we note that conservation laws can impose important constraints on the possible values of scaling exponents.
For example, a Bjorken-expanding medium with a collision integral that conserves particle number (such as \eqref{eq:small-angle-kernel}) satisfies
\begin{align}
\label{n-evo}
  \dot{n}= - 1\,
\end{align}
where the gluon number density is given by $n=\int_{\vp}\, f$. 
In this case, it is easy to show that $\als,\betas,\gas$ must satisfy the relation
\begin{equation}
\label{scaling-relation}
  \als-2\betas-\gas=-1\,  
\end{equation}
even if the exponents themselves are time-dependent.

\section{Scaling in the Fokker-Planck equation}
\label{sec:scaling-num}

Scaling around the BMSS fixed point has been observed before in the FP equation \cite{Tanji:2017suk}. In this section we will establish that this equation also exhibits scaling with a well-defined set of time-dependent scaling exponents prior to approaching the fixed point.

To do this, we numerically solve the Boltzmann equation \eqref{kin} with the collision kernel \eqref{eq:small-angle-kernel} (see Appendix~\ref{sec:numerics} for details on the numerical implementation). 
Following Ref.~\cite{Mazeliauskas:2018yef}, we initialize the gluon distribution at the initial time $\tau_I Q_s = 70$ as
\begin{align}
\label{fI}
 f(\pT,\pz;\tau_I) = \frac{\sigma_0}{g_s^2} \exp \left( - \frac{p_\perp^2 + \xi^2_{0} p_z^2}{Q_s^2} \right)\, , 
\end{align}
where $\xi_{0}$ in eq.~\eqref{fI} characterizes the initial anisotropy and we take $\xi_0 = 2$. 
The parameter $\sigma_{0}$ specifies the overoccupancy of hard gluons at the initial time, i.e. $g^{2}_{s} f(p=Q_{s};\tau_{I})\sim \sigma_{0}$.
For the kinetic description to be valid, we require $\sigma_{0}<1$.

To explore the scaling behavior, we first follow the proposal of Ref.~\cite{Mazeliauskas:2018yef} and study the moments
\begin{equation}
\label{moment-def}
    \text{n}_{m,n}(\tau) \equiv \int_{\vp}  p_\perp^m |p_z|^n f(p_\perp,p_z;\tau)\, , 
\end{equation}
for non-negative integers $m,n$. 
For $m,n<0$, the integration \eqref{moment-def} can potentially be IR divergent.
If the distribution function takes the scaling form~\eqref{prescaling-1}, 
one can substitute this distribution into the definition of moments~\eqref{moment-def}, and find that the percentage change rate of the moments is expressible in terms of scaling exponents as 
\begin{equation}
\label{moment-eq}
    \dot{\text{n}}_{m,n} = \als - (m+2)\betas - (n+1)\gas\, .
\end{equation}

\begin{figure}[t]
\centering
\includegraphics[width=.52\textwidth]{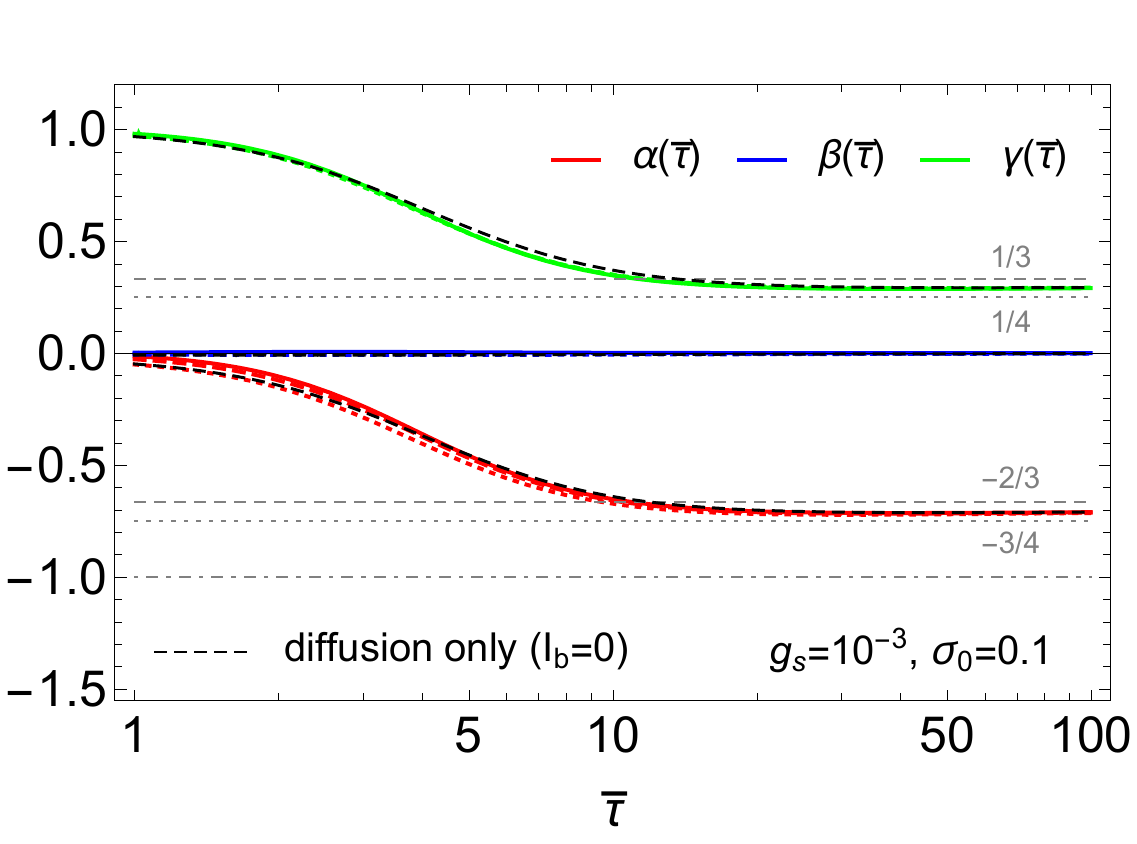}
\includegraphics[width=.52\textwidth]{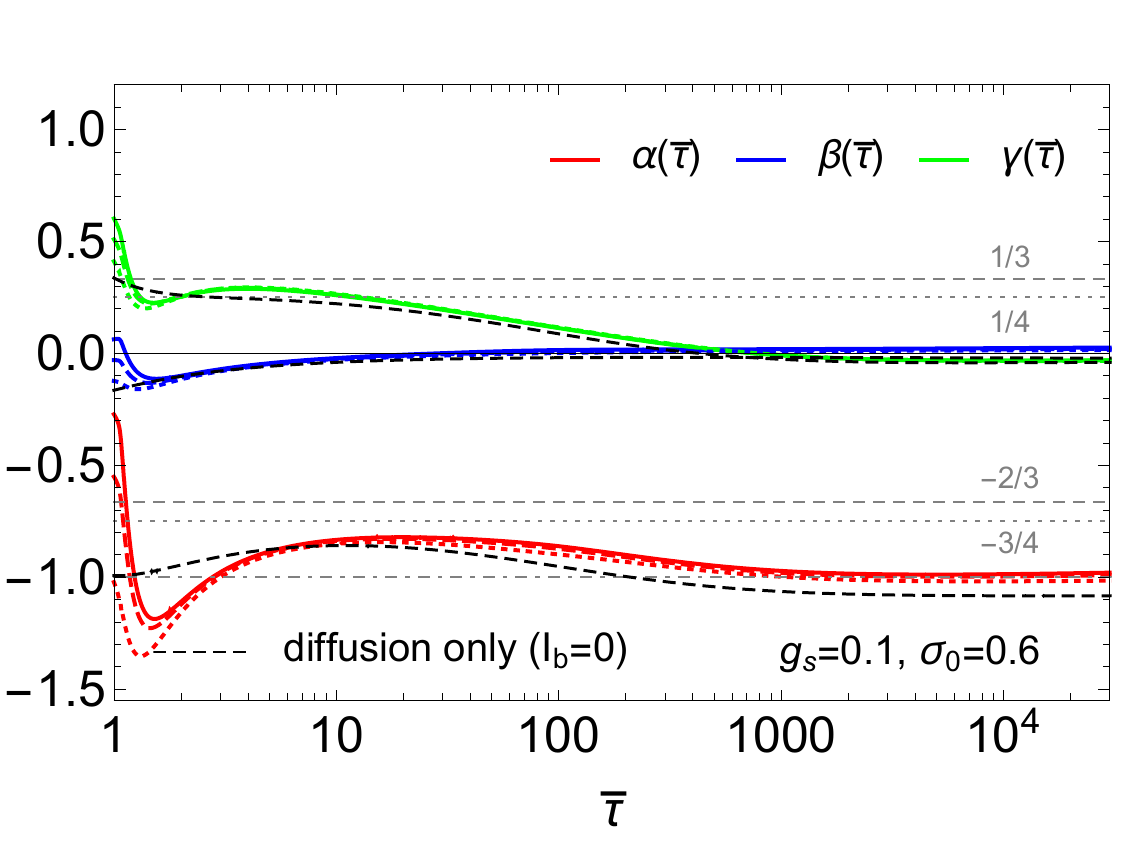}
\includegraphics[width=.52\textwidth]{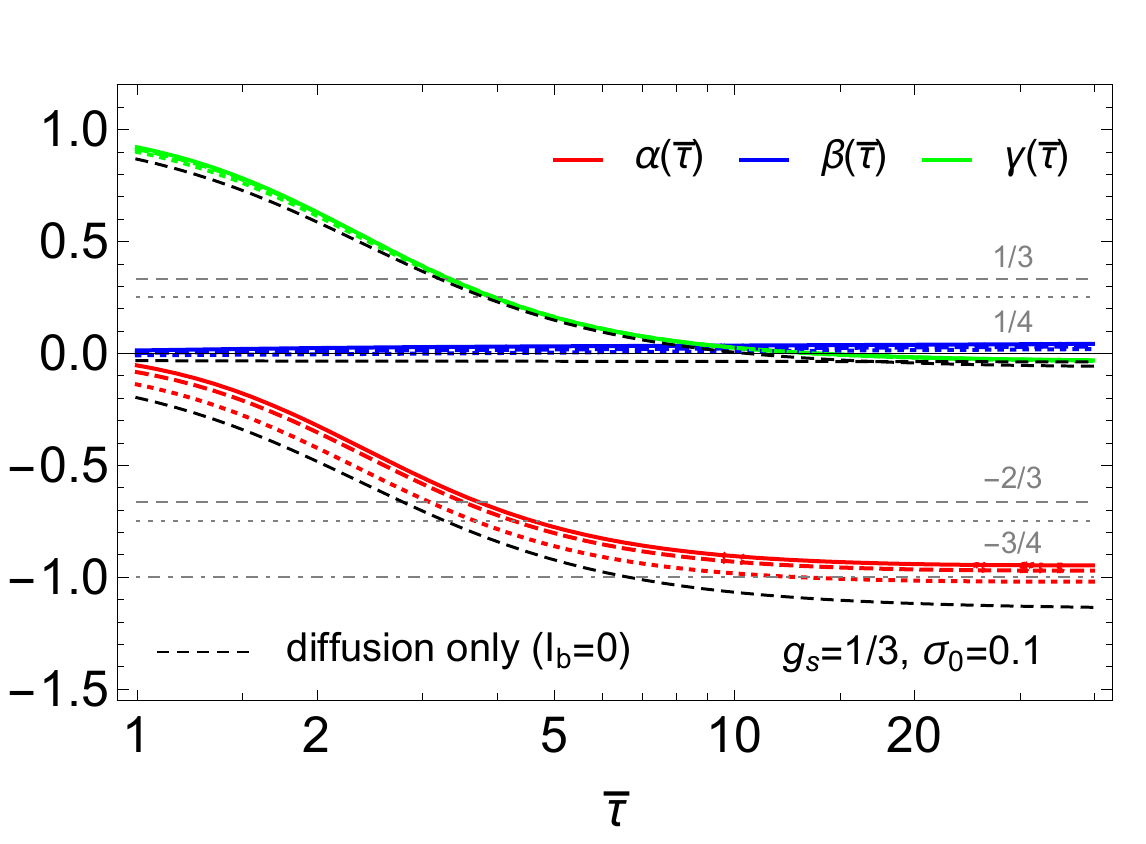}
\caption{
\label{fig:FP-exp}
Evolution of time-dependent scaling exponents $\alpha, \beta, \gamma$ for $(g_s,\sigma_{0})=(10^{-3},0.1)$ (top), $(0.1,0.6)$ (middle) and $(1/3,0.1)$ (bottom) as a function of the rescaled time coordinate $\bar{\tau}=\tau/\tau_{I}$. 
Colored curves show exponents extracted from numerical solutions to the FP equation, with different dashing styles indicating exponents extracted from different combinations of moments in eq.~\eqref{moment-eq}. 
For comparison, black dashed curves show the exponents extracted from solutions with $\I_b=0$ in the collision integral~\eqref{eq:small-angle-kernel} (see text in the three final paragraphs of this section). We include dashed horizontal lines at the values of BMSS and dilute fixed points, and additionally at $1/4$ and $-3/4$ for visual clarity.}
\end{figure}

From numerical solutions to the FP equation we can compute the change rate of any three different moments, and estimate $\alpha(\tau)$, $\beta(\tau)$, and $\gamma(\tau)$ from eq.~\eqref{moment-eq}. 
Throughout we use $\als,\betas,\gas$ to refer to exponents derived by assuming that the distribution has the scaling form, while we use $\alpha,\beta,\g$ for exponents extracted from a general distribution function (as in our numerical results).
In the scaling regime, the exponents extracted from any set of three moments $(m,n)$ via eq.~\eqref{moment-eq} will agree with each other. Conversely, if the system is not in the scaling regime, the exponents extracted from two different sets of moments will generally not agree.
In practice, we obtain the scaling exponents in eq.~\eqref{moment-eq} from three sets of moments 
$(m,n) \in \{(0,0),(1,0), (0,1)\}$;
$(m,n)\in\{(0,0),(2,0), (0,2)\}$; and $(m,n)\in\{(0,1),(2,0), (1,1)\}$.
We take the agreement of exponents extracted from these three sets of three moments as an approximate criterion for the emergence of scaling.

In Fig.~\ref{fig:FP-exp} we present the evolution of the extracted exponents for different combinations of the coupling and initial occupancy, $(g_s,\sigma_0) = (10^{-3},0.1)$, $(0.1,0.6)$, and $(1/3,0.1)$, as a function of the dimensionless time coordinate $\bar{\tau} \equiv \tau/\tau_I$.
The scaling exponents extracted from different sets of moments agree well from $\bar{\tau} \gtrsim 3$ for all $(g_s,\sigma_0)$ combinations, indicating emergence of time-dependent scaling at early times.\footnote{
We note that for $(g_s,\sigma_0)=(0.1,0.6)$ we observe that the relation \eqref{scaling-relation} between the exponents is violated by up to $20\%$ while it is satisfied within a few percent for other $(g_s,\sigma_0)$ combinations shown here.
This effect has also been observed in several previous works on the FP equation \cite{Blaizot:2014jna,Tanji:2017suk} that have suggested it may be related to the gluon condensate at $p=0$. 
}
The late-time behavior of the scaling exponents depends on the combination $(g_{s},\sigma_{0})$. 
For $(g_s,\sigma_{0})=(10^{-3},0.1)$, the late-time values of the exponents are close to (but with visible difference from) the BMSS values $(\als,\betas,\gas)=(-2/3,0,1/3)$. 
This result is also in good agreement with the late-time values of the exponents in QCD effective kinetic theory calculated for the same values of $(g_s,\sigma_0)$ \cite{Mazeliauskas:2018yef}. 
We will discuss in sec.~\ref{sec:a-dim} how this deviation from the exact BMSS scaling exponents can be attributed to an ``anomalous dimension" correction.

Remarkably, in addition to the BMSS fixed point
we also observe a new late-time fixed point in the middle and right panel of Fig.~\eqref{fig:FP-exp}, with (up to small corrections)
\begin{align}
\label{dilute-0}
\textrm{dilute:}\, \qquad
    (\als,\betas,\gas)=(-1,0,0)
\end{align}
We refer to it as the ``dilute" fixed point, since we find that the system evolves to it when the typical occupancy becomes small, $A\ll 1$. 
 For $(g_s,\sigma_0)=(0.1,0.6)$, the exponents tend toward the BMSS values before finally transiting to the dilute fixed point~\eqref{dilute-0}. 
For $(g_s,\sigma_0)=(1/3,0.1)$ the exponents go directly to this new fixed point~\eqref{dilute-0}.
We will further elaborate on its physical origin in sec.~\ref{sec:fixed-point}.

\begin{figure}[t]
  \includegraphics[width=\textwidth]{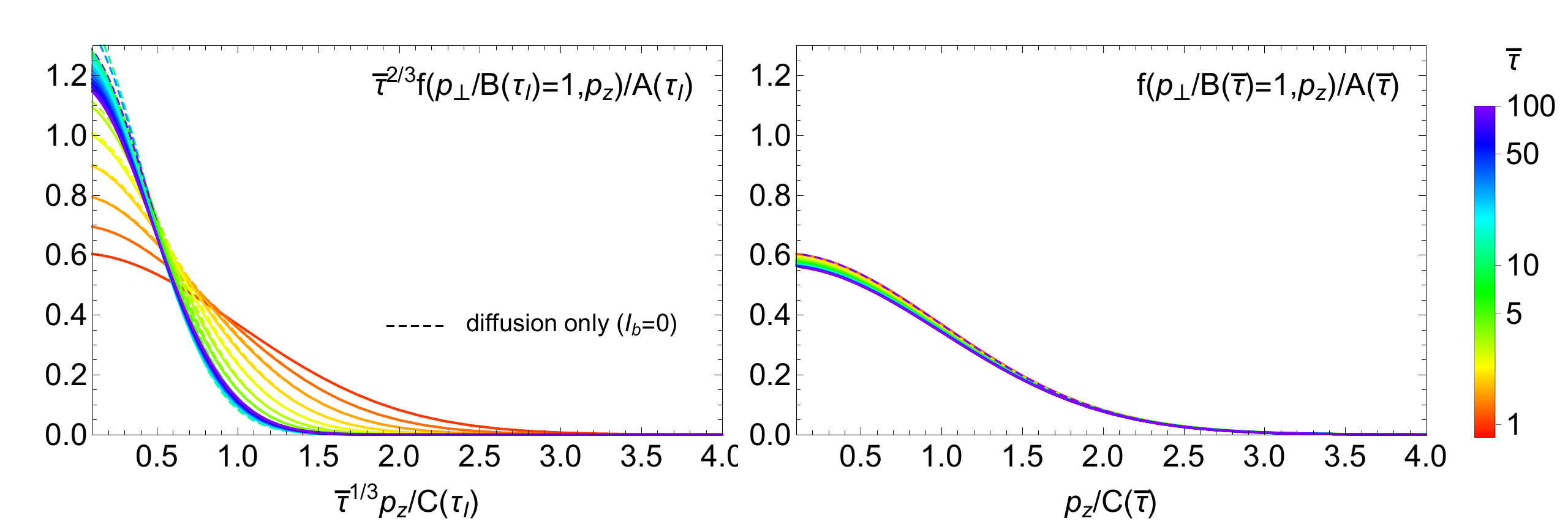}
\includegraphics[width=\textwidth]{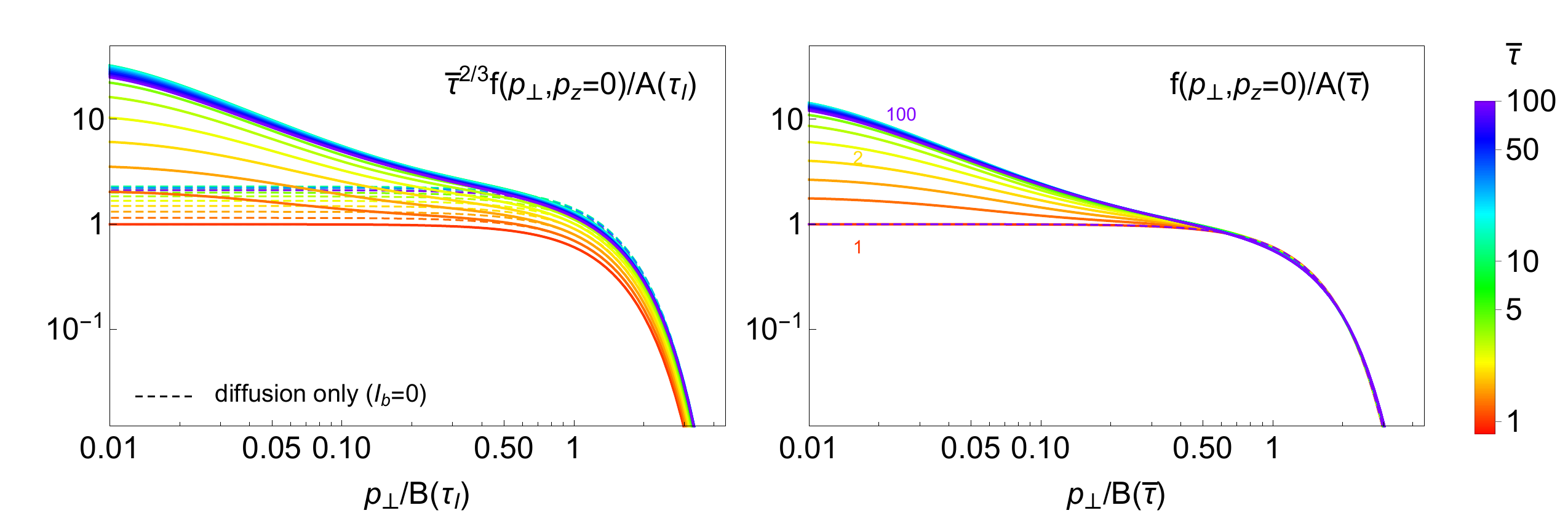}
  \caption{
   \label{fig:FP_case1}
  The rescaled distribution function $w$~\eqref{f-w} for the numerical solutions of the Fokker-Planck equation with $(g_s,\sigma_0)=(10^{-3},0.1)$. 
  The left panel shows the results with $A,B,C$ determined by the BMSS exponent while the right panel shows the same but with time-dependent scaling exponents extracted from Fig.~\ref{fig:FP-exp}. 
  Colors show the evolution in the rescaled time coordinate $\bar{\tau}$. 
  Dashed curves show the analytic scaling solution obtained for $\I_b=0$, i.e., eq.~\eqref{w-sol1}, that has no dependence on $\tau$, as shown later in this work.
}
\end{figure}

\begin{figure}[t]
    \includegraphics[width=\textwidth]{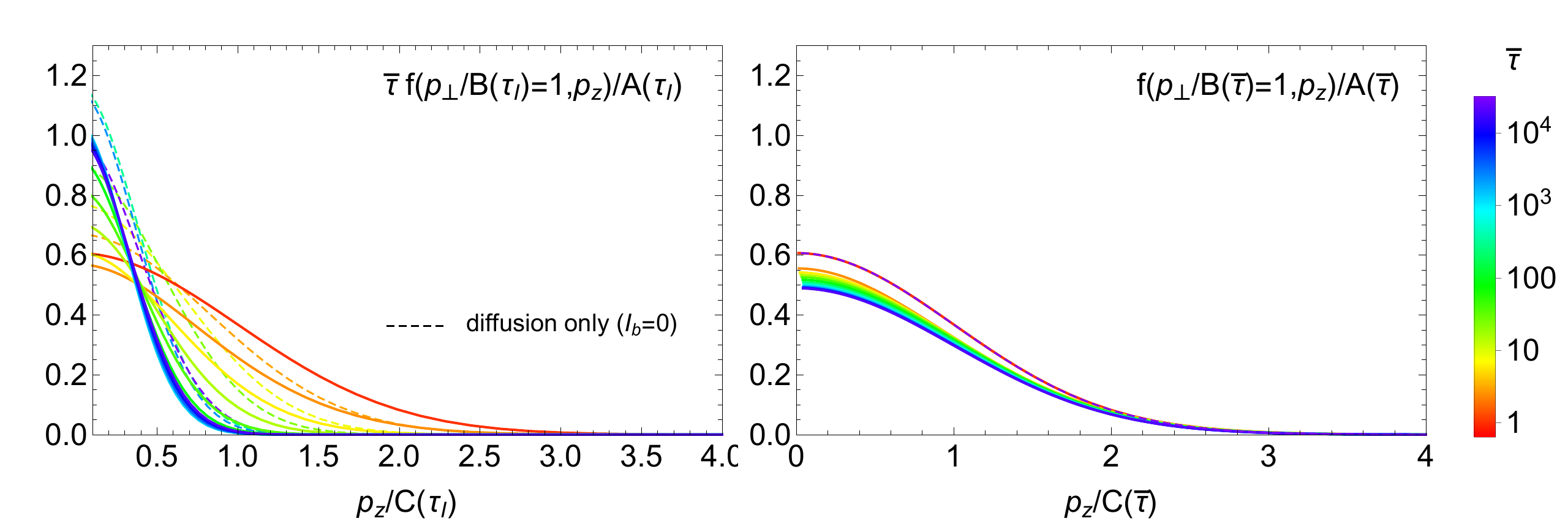}
    \includegraphics[width=\textwidth]{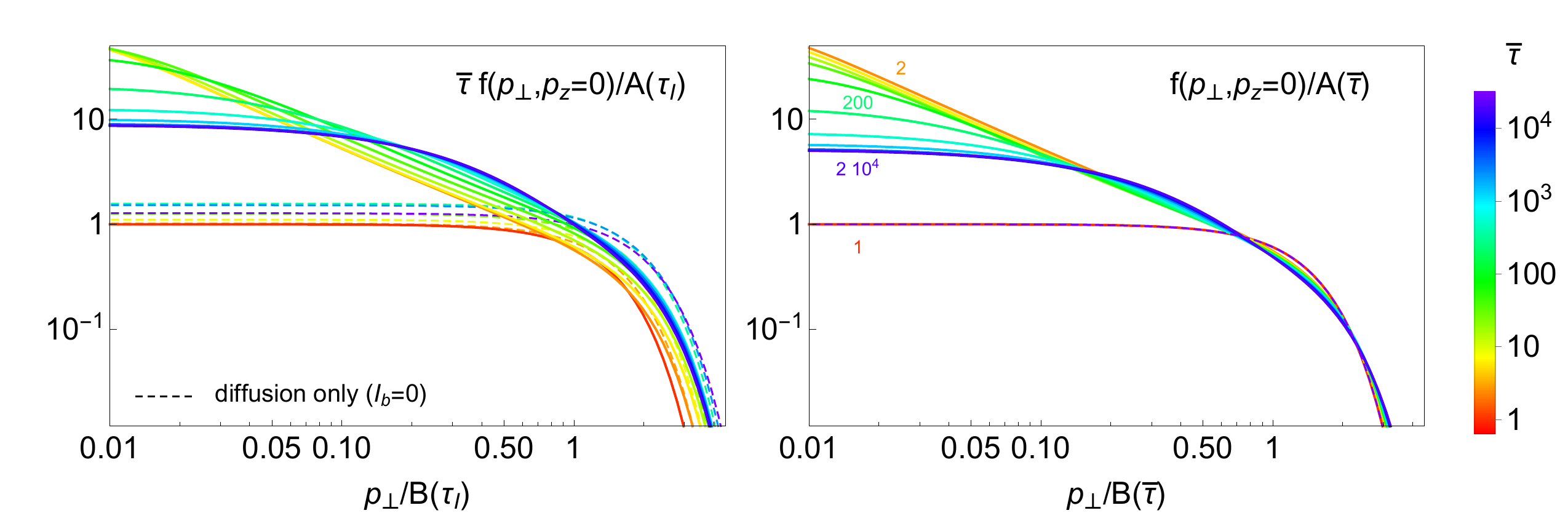}
  \caption{
  \label{fig:FP_case2}
  The same as Fig.~\ref{fig:FP_case1} but for $(g_s,\s_{0})=(0.1,0.6)$. 
  Note in the left figures we use $A,B,C$ determined not by BMSS but dilute fixed point exponents~\eqref{dilute}. 
}
\end{figure}

Though the analysis based on the moment equation~\eqref{moment-eq} shows clearly the evolution of the scaling exponents, the scaling of the distribution is seen more clearly from the full distribution function.
Figs.~\ref{fig:FP_case1} and \ref{fig:FP_case2} show the $\xi$-dependence of the rescaled distribution function $w$ at fixed $\zeta=1$ (top panels) and the $\zeta$-dependence at $\xi=0$ (lower panels) for $(g_s,\sigma_0)=(10^{-3},0.1)$ (Fig.~\eqref{fig:FP_case1}) and $(g_s,\sigma_0)=(0.1,0.6)$ (Fig.~\ref{fig:FP_case2}). We compare the scaling of the distribution function around the fixed point (left panels) to the scaling with time-dependent exponents (right panels). In all panels we take the initial values of $A,B,C$ to be the characteristic occupation number, transverse, and longitudinal momentum of the initial distribution~\eqref{fI}, which gives
$A_{I}=\sigma_{0}/g^{2}_{s}, B_{I}=Q_{s}/\sqrt{2}, C_{I}=Q_{s}/(2\sqrt{2})$. For the left panels we fix (time-independent) exponents $\als,\betas,\gas$ according the late-time fixed point (BMSS in Fig.~\eqref{fig:FP_case1} and dilute in Fig.~\eqref{fig:FP_case2}). In the right panels, we instead estimate the time-dependent scaling exponents $\alpha(\tau)$, $\beta(\tau)$, $\gamma(\tau)$ by averaging the extracted scaling exponents from three sets of moments shown in Fig.~\ref{fig:FP-exp}.

We observe in Fig.~\ref{fig:FP_case1} that, after a short time, the distribution function scales to an excellent degree with time-dependent exponents (right panel) even though the exponents have not yet reached the fixed point.\footnote{To see this from this figure, note that the left and right panels would be equal if the scaling exponents had reached their fixed point values.}
In Fig.~\ref{fig:FP_case2}, scaling appears in the hard regime $\zeta\geq 1$.
Importantly, we see that in both cases scaling occurs before the system reaches the late-time fixed point.  
This agrees with the results shown in Fig.~\ref{fig:FP-exp}. 
We note that the absence of scaling at early times in the soft regime in Figs.~\ref{fig:FP_case1} and \ref{fig:FP_case2} does not contradict our preceding analysis based on the evolution of moments.  
This is because moments with $(m,n)>0$ are mainly determined by the behavior of the distribution in the hard regime $\zeta,\xi>1$, but are less sensitive to that in the soft regime.

%
%

We note that the $\xi$-dependence of the scaling distribution is Gaussian, as is the $\zeta$-dependence in the hard regime.
We shall provide analytic insight into this Gaussianity in sections~\ref{sec:analytic} and~\ref{sec:Adiabatic}.
Collisions among gluons with typical momentum transfer of the order of the Debye mass $m_{D}$~\eqref{mD} will rapidly cascade gluons from the hard regime to the soft regime. 
The growth of the occupancy in the soft regime will in turn expedite the thermalization of soft gluons. 
Therefore, we observe $1/\zeta$ behavior in Fig.~\ref{fig:FP_case1}, corresponding to the small momentum limit of Bose-Einstein distribution. 
For Fig.~\ref{fig:FP_case2}, the system transits from the BMSS fixed point to the dilute fixed point (see Fig.~\ref{fig:FP-exp} (middle)), meaning the typical occupancy decreases from $A\gg 1$ to $A\ll 1$. 
Accordingly, the distribution at small $\zeta$ first shows $1/\zeta$ behavior and then becomes more similar to a Boltzmann distribution at later times.

As we demonstrate in Appendix~\ref{sec:Ib}, the $\I_{b}$ term is less important than the $\I_{a}$ term
in the hard regime and when $A\geq 1$. 
To verify this in our numerical approach, we also compute the scaling exponents for a purely diffusive kernel by setting $\I_b=0$ in eq.~\eqref{eq:small-angle-kernel}. 
The resulting exponents are shown in dashed black lines in Fig.~\ref{fig:FP-exp}. We observe good agreement between the exponents obtained from solving the full FP equation and those obtained with setting $\I_{b}=0$. We also show the scaling function for $\I_{b}=0$ in dashed lines in Figs.~\ref{fig:FP_case1} and ~\ref{fig:FP_case2}. 
The scaling distribution with $\I_{b}=0$ describes that of the hard gluons very well, in particular when $\As$ is not too small. We will show in sec.~\ref{sec:analytic} that 
the scaling function $w_{S}$ can also be computed analytically for $\I_{b}=0$, see eq.~\eqref{w-sol1}.

In summary, we have observed time-dependent scaling behavior in the FP equation for hard gluons $\zeta=p_{\perp} \Bs\geq 1$.
We find that the FP equation captures the key qualitative behavior of time-dependent scaling in EKT in this hard regime for $(g_s,\sigma_0)=(10^{-3},0.1)$, as was first shown in Ref.~\cite{Mazeliauskas:2018yef}.
In contrast, we do not observe early time scaling for soft gluons, indicating the importance of inelastic scattering in the soft regime (as was already noticed in Ref.~\cite{Mazeliauskas:2018yef}).
Nevertheless, 
our finding suggests that scaling of hard gluons is mainly driven by the longitudinal expansion and  $2\leftrightarrow 2$ small angle scatterings that are present in the FP equation. 
When the system is not too dilute, 
the solution to the FP equation in the hard regime is well-described by considering only the diffusive term (proportional to $\I_{a}$) in the collision integral. Of course, the $\I_{b}$ term is important in the soft regime where gluons are in equilibrium, since the equilibrium distribution in this regime is crucially determined by the balance between $\I_{a}$ and $\I_{b}$ terms.

In the coming sections we will understand the emergence of scaling in an analytically transparent way for hard gluons, by studying the FP equation with $\I_{b}=0$. We will do so incrementally. First, in sec.~\ref{sec:analytic} we will study the solutions that exhibit scaling considering only the longitudinal part of the collision kernel~\eqref{CIa-0}. Second, in sec.~\ref{sec:Adiabatic} we will demonstrate the relevance of adiabatic evolution in this problem, and explain why the self-similar solutions are dynamically preferred. Finally, in sec.~\ref{sec:evo} we will derive the evolution equations for the time-dependent scaling exponents and compare with the numerical solutions to the FP equation as well as the results from EKT simulations in Ref.~\cite{Mazeliauskas:2018yef}.

\section{Analytic scaling solution for longitudinal diffusion
\label{sec:analytic}
}

In this section, 
we shall derive the scaling solution to the FP equation analytically in the limit that the typical longitudinal momentum is much smaller than the typical transverse momentum, i.e. $C/B \ll 1$. 
To leading order in small $C/B$,
the collision integral~\eqref{eq:small-angle-kernel} is reduced to eq.~\eqref{CIa-0}, and we can write the FP equation as
\begin{align}
\label{f-simple}
\pd_{y}f=\le(p_z \pd_{p_z} +q\, \pd_{p_z}^2\ri)\, f\, .
\end{align}
Here, we have defined the effective momentum diffusive constant
\begin{equation}
\label{q-def}
  q \equiv \lambda_0 \lcb\, \I_a[f]\, \tau \, ,
\end{equation}
which, for later convenience, is defined with respect to a (dimensionless) logarithmic temporal variable
\begin{align}
y\equiv \log(\tau/\tau_{I})\, . 
\end{align} 
Though $q$ is a functional of $f$, for notational brevity we leave this dependence implicit.
Since the simplified collision integral~\eqref{CIa-0} does not change transverse momentum, in this section
we shall suppress the $\pT$-dependence in the distribution function and set $\betas=0$.

To look for a self-similar solution, 
we substitute eq.~\eqref{prescaling-1} into eq.~\eqref{f-simple} to obtain an equation for $w_{S}$:
\begin{align}
\label{w-eq-full}
  \pd_{y}w_{S}&=-\als w_{S} + (1-\gas)\xi\,\pd_{\xi}w_{S}+\frac{\qs}{C^{2}_{S}}\pd^{2}_{\xi}w_{S}
\nonumber  \\
  &=
  - (1-\gas)\, 
  \le[
  \frac{\qs}{(1-\gas)\Cs^{2}}\pd^{2}_{\xi}+\xi\,\pd_{\xi}-\frac{\als}{(1-\gas)}
  \ri] w_S
\end{align}
where we have used the definitions of $\als,\gas$ from~\eqref{exp-def}. 
Here the scaling variable $\xi$~\eqref{zeta-xi-def} is evaluated with $C=\Cs$ and the subscript ``$S$" in $\qs$ reminds us that $q$ is evaluated with the scaling distribution function as its argument. 

Then, by definition, a distribution undergoing scaling satisfies $\partial_y w_S = 0$, and consequently eq.~\eqref{w-eq-full} becomes
\begin{align}
\label{ws}
 w_{S}+\xi\,\pd_{\xi}\,w_{S}+ \frac{\qs}{(1-\gas)\Cs^{2}}\pd^{2}_{\xi}\,w_{S}=0\, , 
\end{align}
where we have used the relation among scaling exponents~\eqref{scaling-relation} with $\betas=0$, namely $\als=-1+\gas$.
In the analysis above, we have assumed $\gas\neq 1$.
In the special case $\gas=1$, the condition on the scaling solution can be read from the first line of eq.~\eqref{w-eq-full}: $\qs\,\pd^{2}_{\xi}w_{S}=0$, which has no bounded solution unless $\qs=0$. 
The latter corresponds to the free-streaming (collisionless) limit with scaling exponents given by $(\als,\betas,\gas)=(0,0,1)$.

For $w_{S}$ determined by eq.~\eqref{ws} to be time-independent, we must have
\begin{align}
  \frac{(1-\gas)}{\qs/\Cs^{2}}=\textrm{const}\, . 
\end{align}
Without loss of generality, we choose the normalization of $\Cs$ such that
\begin{align}
\label{C-cond}
  \frac{(1-\gas)}{\qs/\Cs^{2}}=1\, ,
\end{align}
and with this choice, we can write the equation for $w_{S}$~\eqref{ws} as
\begin{align}
\label{ws-0}
   &\, \pd^{2}_{\xi}\,w_{S}+\xi\,\pd_{\xi} w_{S} + w_{S} = 0\, . 
\end{align}
Note that eq.~\eqref{C-cond} imposes a non-trivial self-consistency condition for $\Cs$ since $\qs$ itself is a functional of the distribution function. 
Moreover, because $\gamma_S = -\dot{C}_S$, this equation is also implicitly a differential equation for $\Cs$.
Therefore, the evolution of $\Cs$, and consequently $\gas$, can be determined by solving eq.~\eqref{C-cond} (see sec.~\ref{sec:evo} for more details). 
Up to a normalization constant, the solution to eq.~\eqref{ws-0} is
\begin{align}
\label{w-sol}
  \lim_{\Cs/\Bs\to 0}\, w_{S}(\zeta,\xi)=  e^{-\frac{\xi^2}{2}}\, .
\end{align} 
The other linearly independent solution to the differential equation~\eqref{ws-0} does not give a finite number density when integrated over the momentum domain $\xi$, and hence has to be discarded.

As we have noted in sec.~\ref{sec:kin-setup}, when the typical occupancy is large
the first order corrections from $\Cs/\Bs$ can be accounted for by setting $\I_{b}=0$ in the FP collision integral~\eqref{eq:small-angle-kernel}, but keeping the derivatives with respect to $p_{\perp}$ in the $\I_{a}$ term. 
In this case, the collision integral is reduced to eq.~\eqref{CIa-1}. 
By a straightforward generalization of the analysis presented in this section, we find that the scaling solution is given (up to normalization) by
\begin{align}
\label{w-sol1}
    w_{S}(\zeta,\xi)=  e^{-\frac{\zeta^{2}+\xi^2}{2}}\, . 
\end{align}
Eq.~\eqref{w-sol} and its generalization eq.~\eqref{w-sol1} are the main results of this section. 
They tell us that the momentum dependence is Gaussian in the scaling regime, which is also what we observed numerically in the previous section. 
In the next section, we shall explain why the distribution function is attracted to this scaling form, using the adiabatic theorem of quantum mechanics as our main guiding principle.

\section{Scaling and adiabaticity} 
\label{sec:Adiabatic}

Here we set out to demonstrate the close connection between the emergence of scaling behavior and adiabaticity in the temporal evolution of the distribution function. 
Let us first recall that in a time-dependent quantum mechanical problem where the Hamiltonian changes with time, a system prepared in its ground state will remain in the instantaneous ground state as long as the transition rates between the ground and excited states are small compared to the energy gap between them. This is referred to as adiabatic evolution and characterizes many real-time dynamical problems in quantum mechanics~\cite{APT}.

In Ref.~\cite{Brewer:2019oha},
the idea of adiabatic evolution has been employed to describe the far-from-equilibrium evolution of the Boltzmann equation for a Bjorken-expanding plasma under the relaxation time approximation 
(see also Ref.~\cite{Blaizot:2021cdv}). With a natural, yet non-trivial, extension, 
we shall see that the scaling evolution obtained in the previous section can be viewed as an example of adiabatic evolution. 
In particular, we will show that adiabaticity naturally explains why the rescaled distribution function $w$ will generically be attracted to and stay in a time-dependent scaling function  $w_{S}$.

\subsection{Adiabatic frame
\label{sec:adi-pz}
}

For definiteness, we shall begin with the simplified collision integral~\eqref{CIa-0} and suppress the $\pT$ dependence of the distribution function.
In the next subsection we will extend our analysis to the collision integral defined by~\eqref{CIa-1}.

To make contact with quantum mechanics, we recast the evolution equation~\eqref{w-eq-full} for the rescaled distribution function $w$ into the form
\begin{equation} 
\label{w-eq-H}
    \partial_y w = - {\cal H}\, w\, ,
\end{equation}
where the ``Hamiltonian" operator reads
\begin{equation}
    \label{H}
    {\cal H} = - (1-\g)\, 
  \le[
  \tq\, \pd^{2}_{\xi}+\xi\,\pd_{\xi}-\frac{\alpha}{(1-\gamma)}
  \ri]\, ,
\end{equation}
and we have defined
\begin{equation}
\label{tq-def}
  \tq = \frac{q}{C^{2}(1-\g)}\, .
\end{equation}
Note that $\tq$ is a functional of $A,C$ since $q$ (defined in eq.~\eqref{q-def}) depends on the distribution function $f$ and hence in general is evolving in time. 
Eq.~\eqref{w-eq-H} is analogous to the Schr\"odinger equation except that the operator ${\cal H}$ is non-Hermitian because the system under study is expanding and involves dissipative processes due to collisions.

Since $\mathcal{H}$ is a non-Hermitian operator, its left and right eigenvectors are not necessarily related to each other by complex conjugate. We have
\bes
\begin{align}
  {\cal H}(y) \phi_{n}^R(\xi;y)&= \sE_{n}(y)\phi_{n}^R(\xi;y)\, 
\\
  {\cal H}^\dagger(y) \phi_{n}^L(\xi;y)&= \sE_{n}(y)\phi_{n}^L(\xi;y)\, 
\end{align}
\ees
where the conjugate of ${\cal H}$ is given by
\begin{equation}
    {\cal H}^\dagger w = -(1-\gamma) \left[\tq \partial_{\xi}^2 w - \partial_{\xi} ( \xi w)  - \frac{\alpha}{1-\gamma} w \right]\, .
\end{equation}
The eigenfunctions of ${\cal H}$ represent a specific form of the distribution function in phase space, and as such, must have finite support in $\xi$ space.
Furthermore, assuming inversion symmetry about the longitudinal axis $p_z \to -p_z$, the eigenfunctions should be even in $\xi$.
Taking these constraints into account, we find 
\begin{align}
\label{phi}
&\, \phi_{n}^L = 
{\rm He}_{2n} \! \left(\frac{\xi}{\sqrt{\tq}} \right) \, ,
\qquad
  \phi_{n}^R= 
\frac{1}{\sqrt{2\pi \tq}(2n)!}\, {\rm He}_{2n}\! \left(\frac{\xi}{\sqrt{\tq}} \right)\, e^{-\frac{\xi^{2}}{2\tq}}\, ,
\\
\label{En}
&\,\sE_{n} = 2n(1-\g) +(\a-\g+1)\, , 
\end{align}
for $n=0,1,\ldots$.
Here ${\rm He}_{2n}$ denote probabilist's Hermite polynomials, 
and we have chosen the normalization of eigenstates by requiring $\int_{-\infty}^\infty d\xi \, \phi_m^L(\xi) \, \phi_n^R(\xi) = \delta_{mn}$. For obvious reasons, we refer to the $n=0$ mode as the instantaneous ground state
\begin{align}
\label{gs}
  \phi_{0}^R(\xi;y)=\frac{1}{\sqrt{2\pi \tq}}\,e^{-\frac{\xi^2}{2\tq}}\,  
\end{align}
and to modes with $n>0$ as instantaneous excited states.

The defining property of a system undergoing adiabatic evolution is that the contribution from excited states through transitions is suppressed.
To quantify the weight of excited states in the rescaled distribution $w$, we write
\begin{align}
w(\xi;y)=\sum_{n=0}\, \ta_{n}(y)\phi_{n}^R(\xi;y)\, .
\end{align}
From eqs.~\eqref{w-eq-H},~\eqref{H}, and the orthogonality of the eigenbasis, it can be shown that the coefficients $\ta_n(y)$ follow the evolution equation~(see also Ref.~\cite{Brewer:2019oha})
\begin{equation}
\label{an}
    \partial_y \ta_n + \sum_{m \neq n}  V_{nm}(y) \ta_m = -\sE_n(y) \ta_n\, ,
\end{equation}
with
\begin{equation}
\label{Vmn}
    V_{mn} = \int^{\infty}_{-\infty} d\xi \, \phi_m^L(\xi;y) \, \partial_y \phi_n^R(\xi;y)=  \dot{\tq}\,  m(2m-1) \delta_{m-1,n}\, .
\end{equation}
Transitions between different eigenstates occur only through $V_{mn}$ in eq.~\eqref{an}. Since the eigenstates~\eqref{phi} depend on time through the time-dependence of $\tq$, 
the transition rate~\eqref{Vmn} is proportional to $\pd_{y}\tq$. When this transition rate is not small, the ground state can mix with excited states and adiabaticity will break down.

However, at this point we have the freedom to choose $A,C$ at will, so we will look for $A,C$ that minimize the transition rate $V_{mn}$. 
This goal can be achieved by imposing the condition
\begin{align}
\label{q-choice}
\tq = \frac{q}{C^{2}(1-\g)}=1\,  ,
\end{align}
so that $V_{mn}$~\eqref{Vmn} vanishes. 
As in sec.~\ref{sec:analytic}, 
we shall assume $\g<1$, so that the ground state is gapped from the excited states $\phi_{n>0}$ by $2n(1-\g)$. With $V_{nm}=0$ and an energy gap between the ground and excited states, the conditions for adiabaticity are satisfied.
With~\eqref{q-choice}, the eigenfunctions~\eqref{phi} do not depend on time explicitly and become
\begin{align}
\label{phi-2}
   \phi_{n}^R(\xi)&= \frac{1}{\sqrt{2\pi}(2n)!}\, \text{He}_{2n}(\xi)\, e^{-\frac{\xi^{2}}{2}}\, , 
\end{align}
Up to now, we still have the freedom to specify $\a$. 
A natural choice is to take
\begin{align}
\label{alpha-choice}
\alpha = \gamma -1\,      
\end{align}
such that the ground state energy $\sE_{0}$ in eq.~\eqref{En} vanishes,
\begin{align}
    \sE_{0}=0\, . 
\end{align}
We will define the frame satisfying conditions~\eqref{q-choice},~\eqref{alpha-choice} as the ``adiabaticity frame,'' and denote the associated rescalings by $A_{\ad},C_{\ad}$.

Since we have seen that the conditions for adiabaticity are satisfied in the adiabatic frame, we expect that $w$ will approach the ground state following the decay of excited states,
\begin{align}
\label{gs-co}
   w\to \phi_{0}^R&= \frac{1}{\sqrt{2\pi}}\, e^{-\frac{\xi^{2}}{2}}\, ,
\end{align}
and then will remain in the ground state because the evolution is adiabatic. 
In this case, $\pd_{y}w=-\sE_{0}w=0$ by construction. 
Therefore, we can identify $\phi_{0}$ with the scaling distribution $w_{S}$. This identification can also be confirmed by looking at the explicit expression for $w_{S}$~\eqref{w-sol}. 
We also note that the conditions~\eqref{C-cond},~\eqref{scaling-relation} determining  $\As,\Cs$ are the same as those specifying the adiabatic frame~\eqref{q-choice},~\eqref{alpha-choice}.
We therefore conclude that during the adiabatic evolution $A_{\ad},C_{\ad}$ and $\As,\Cs$ coincide and that scaling behavior for the collision integral~\eqref{CIa-0} is an example of adiabatic evolution.

We wish to emphasize the similarities and differences between $\As, \Cs$ and $A_{\ad},C_{\ad}$. 
Only in the scaling regime is it possible to identify $\As,\Cs$ such that the rescaled distribution $w$ becomes time-independent.
On the other hand, the adiabatic frame $A_{\ad},C_{\ad}$ is defined by requiring the evolution of the instantaneous eigenstates of ${\cal H}$ to be as slow as possible. 
Such a frame exists even if $w$ is different from $w_S$. 
Indeed, for a given $q$, the corresponding $C_{\ad}$ can be obtained by solving \eqref{q-choice}, without having to require that the system is in the ground state.

We finally note that one could study the dynamics of the distribution function in different frames, and still solve the same physical problem. 
The advantage of using the adiabatic frame is that this frame reveals the adiabatic nature of the scaling evolution.  
Moreover, in this frame we can conveniently describe how a self-similar evolution for the distribution function arises from a generic initial condition: the ground state, i.e., the scaling distribution, becomes the dominant contribution to the state of the system through the decay of excited states. Explicitly, since $V_{nm}=0$ in the adiabatic frame, the evolution equation for $\ta_n$~\eqref{an} reads
\begin{align}
\label{a-evo}
  \pd_{y} a_{n}=
  -\sE_{n}\, a_{n}=
  - 2n (1-\g) \,a_{n}\, ,
\end{align}
from which it is clear that the excited modes decay as $\sim e^{-2(1-\gamma)ny}$, and only the ground state survives after a transient time. 
This is why the scaling distribution is an attractor of the evolution.

\subsection{Time scales for approaching the scaling function and approaching the fixed point} 
\label{sec:time-compare}

Time-dependent scaling, such as that observed in Ref.~\cite{Mazeliauskas:2018yef}, occurs when the time scale for the system to approach the scaling distribution $\tau_S$ is much shorter than that for the exponents to reach their fixed point values $\tau_\text{FP}$.
In this section we wish to understand the condition under which $\tau_S \ll \tau_\text{FP}$.
In this situation, 
the evolution of the distribution function is captured by the evolution of the scaling exponents from $\tau_S$ to $\tau_\text{FP}$.

The analysis in the previous section tells us that distribution will approach the scaling form (ground state) after the damping of excited states. 
Therefore, $\tau_{S}$ is set by the inverse of the energy gap between the ground and excited states in the adiabaticity frame. 
For illustrative purposes we can estimate $\tau_{S}$ from eq.~\eqref{a-evo}, assuming that we are not very close to the free-streaming limit so that $\gamma$ in the adiabatic frame is not very close to $1$. 
Under this condition, 
the time scale for the decay of the $n^\text{th}$ excited state is $\tau_{I} \exp(\frac{3}{4n})$.
On the other hand, 
eq.~\eqref{g-analytic} (derived in sec.~\eqref{sec:evo}) 
tells us that deviations from the BMSS fixed point value $\g_{{\rm BMSS}}=1/3$ will decay as $e^{-2y}$, meaning that $y_{\rm FP} \sim 1/2$ or $\tau_{{\rm FP}} \sim \tau_{I}\, \sqrt{e}$.
Therefore, the excited states with sufficiently large $n$ decay at a scale much shorter than $\tau_{\rm FP}$. 
However, low lying excited states decay on a similar time scale to the approach to the fixed point and therefore may defer the emergence of scaling and shorten, or even remove altogether the time-dependent scaling regime.  
A long duration of time-dependent scaling requires a clean separation $\tau_{S}\ll \tau_{{\rm FP}}$, which requires that the contribution from the low-lying excited states in the initial distribution be sufficiently small. 
This is consistent with our numerical observation in sec.~\ref{sec:scaling-num} that with a Gaussian initial condition, scaling starts at early times. Gaussian initial conditions were also used for solving QCD EKT in Ref.~\cite{Mazeliauskas:2018yef}.

\subsubsection{Contribution from the first excited state}

Our previous observation notwithstanding, even when the contribution from the first excited state is significant, one can still show that a time-dependent scaling phase exists prior to reaching the fixed point values. Consider a distribution function given by
\begin{equation}
    w = a_0 \phi^{R}_{0} + a_1 \phi^{R}_{1} = \frac{a_0}{\sqrt{2\pi}} e^{-\frac{\xi^2}{2}} \left[ 1 + \delta \frac{\xi^2 - 1}{2} \right] \, ,
\end{equation}
where we have introduced $\delta \equiv a_1/a_0$, representing the relative contribution of the first excited state to the full distribution function.

When $\delta > 1$, there is no reason to expect that any kind of self-similarity will appear in the distribution function. However, if $\delta < 1$, one can manipulate the previous expression into
\begin{equation}
    w = \frac{a_0}{\sqrt{2\pi (1+\delta)}} e^{-\frac{\xi^2}{2(1+\delta)}} + O(\delta^2) \, ,
\end{equation}
from which it is apparent that the full distribution function $f(p_z; \tau) = A\, w(p_z/C ; \tau )$ has a scaling form (at least when $\delta$ is perturbatively small), albeit with a different set of rescaling functions:
\begin{align}
    A &\to A_{\delta} = \frac{A}{\sqrt{1+\delta}} \, , \\
    C &\to C_{\delta} = C \sqrt{1+\delta} \, .
\end{align}
Then, by using that $\partial_y \delta = \partial_y \ta_1 / \ta_0 = - 2(1-\gamma) \delta$, one immediately infers that the distribution function $f$ will exhibit scaling, with exponents given by
\begin{align}
    \alpha_\delta &= \alpha + \delta (1 - \gamma) + O(\delta^2) \, , \\
    \gamma_\delta &= \gamma + \delta (1 - \gamma) + O(\delta^2) \, .
\end{align}
What is perhaps most remarkable about this is that a precise notion of scaling survives up to linear order in $\delta$, which expands the domain of time-dependent scaling phenomena even further. This result guarantees that, at the very least, there will always be a short time-dependent scaling phase once $\delta$ becomes sufficiently small before reaching the attractor.

\subsection{
Generalization to isotropic diffusive kernel
\label{sec:with-pT}
}

To complete our discussion on adiabaticity for the FP equation, in this section we will extend the adiabatic analysis in sec.~\eqref{sec:adi-pz} to the collision integral~\eqref{CIa-1}, which includes transverse momentum diffusion. 
The evolution equation for $f$ is now
\begin{equation}
\label{FP_diffusive}
    \pd_{y}f=\le(p_z \pd_{p_z} +q \nabla^{2}_{\bf p}\ri) f\, . 
\end{equation}
Since this equation describes diffusion in transverse momentum, we shall reinstate the $p_T$-dependence of the distribution function.
 From the definition of the scaled distribution function $w$~\eqref{f-w}, eq.~\eqref{FP_diffusive} can be rewritten as $\partial_y w = - {\cal H} w$, with
\begin{equation}
    \label{H-iso}
    {\cal H} = \alpha - (1-\g) 
  \le[
  \tq\, \pd^{2}_{\xi}+\xi\,\pd_{\xi}
  \ri] + \b  
  \le[
   - \frac{q}{B^2 \b}\,( \pd^{2}_{\zeta} +  \frac{1}{\zeta} \pd_{\zeta}) + \zeta\,\pd_{\zeta}
  \ri]\, ,
\end{equation}
where $\tq$ is defined in eq.~\eqref{tq-def}.

Analogously to our analysis in sec.~\ref{sec:adi-pz}, 
we choose $A,B,C$ to ensure the adiabatic evolution of the states in this system. 
It is straightforward to show that the appropriate choice is 
\begin{align}
\label{q-general}
   \tq = \frac{q}{C^2 (1-\gamma)}=1\, ,
   \qquad
\tqB \equiv - \frac{q}{B^2 \b}=1 \, .
\end{align}
Furthermore, imposing the condition
\begin{align}
\label{alpha-general}
\alpha = \gamma + 2\beta - 1 \, ,
\end{align}
we can make the ground state energy zero (note that when the distribution is in the scaling regime, this is implied by number conservation).
In this adiabatic frame the eigenvalues of ${\cal H}$ are
\begin{equation}
\label{eigen-iso}
    \sE_{n,m} = 2n (1-\gamma) - 2m \beta \quad \quad n,m = 0,1,2,\ldots \, ,
\end{equation}
which can be verified explicitly by solving for the eigenfunctions of each operator in the square brackets of~\eqref{H-iso}. For the longitudinal part, they are Hermite functions as before, whereas for the transverse part they are given by confluent Hypergeometric functions: 
\begin{align}
    &\, \phi_{n,m}^L = 
{\rm He}_{2n} \! \left(\frac{\xi}{\sqrt{\tq}} \right) {}_1 F_1 \! \left(-2m,1,\frac{\zeta^2}{2 \tqB }\right) \, , \\
  & \, \phi_{n,m}^R= 
\frac{1}{\sqrt{2\pi \tq} (2n)!} \frac{1}{\tqB} \, {\rm He}_{2n}\! \left(\frac{\xi}{\sqrt{\tq}} \right)  {}_1 F_1 \! \left(-2m,1,\frac{\zeta^2}{2 \tqB }\right) e^{-\frac{\xi^{2}}{2\tq}-\frac{\zeta^{2}}{2\tqB}} \, \, .
\end{align}
One can verify that these Hypergeometric functions are actually polynomials and that the states are normalized under the inner product 
\begin{equation}
    \int_{-\infty}^\infty \!\!\!\! d\xi \int_0^\infty \!\!\!\! d\zeta \, \zeta \, \phi^L_{n_L,m_L}(\xi,\zeta) \phi^{R}_{n_R,m_R}(\xi,\zeta) = \delta_{n_L, n_R} \delta_{m_L,m_R} \, .
\end{equation}
For the reasons listed below eq.~\eqref{BC-ratio}, and in consistency with~\eqref{q-general}, 
we have assumed $\beta\leq 0$. 
With the choice $\tq = \tqB = 1$, the ground state $(n,m)=(0,0)$ is given exactly by
\begin{equation}
\phi_{0,0}^R = \frac{1}{\sqrt{2\pi}} e^{-\frac{\xi^2+\zeta^2}{2}} \, ,
\end{equation}
which coincides with the scaling solution~\eqref{w-sol1} of the same collision integral. 
This again illustrates the connection between adiabaticity and scaling evolution.

Since $\beta$ is assumed to be small, we note that 
the energy gap  $-2 m \beta$ between the ground state $\phi^{R}_{0,0}$ and ``transverse'' excited states $\phi^{R}_{0,m}$ is not particularly large. 
This means that in general, the longitudinal profile approaches the Gaussian form much earlier than the transverse profile does. 
Physically, this difference means that the longitudinal expansion changes the longitudinal momentum distribution rather rapidly. 
Applying the argument presented in sec.~\ref{sec:time-compare}, we conclude that for the transverse profile to exhibit scaling with a universal distribution form $w_S$ prior to approaching the fixed point, the initial transverse distribution should be close to a Gaussian, because deviations from Gaussianity (i.e., from excited states) would typically be long-lived.

\section{The evolution of scaling exponents
\label{sec:evo}
}

An important implication of scaling is that it simplifies the description of the gluon plasma evolution far from equilibrium. 
Once the scaling function is given, the evolution in the scaling regime can be described by that of time-dependent scaling exponents $\als,\betas,\gas$. In this section we derive evolution equations for scaling exponents from the same conditions that ensure adiabaticity for the collision kernel~\eqref{CIa-1}.
As shown below, the resulting equations lead to various fixed points, and provide a precise description of the flow between those fixed points.

\subsection{Deriving evolution equations
\label{sec:evo-eq}
}

In the previous section, we have demonstrated that one can choose $A,B,C$ (and consequently $\a,\b,\g$) such that the scaling distribution $w_{S}$ is the ground state of the Hamiltonian ${\cal H}$ that describes the evolution of $w$ with zero eigenvalue. 
This leads to the self-consistency conditions~\eqref{q-general}, \eqref{alpha-general}, which in the scaling regime become
\begin{align}
\label{alpha-relation}
\als &=-2\betas-\gas-1
\\
  \label{beta-q}
  -\betas &= \frac{\qs}{\Bs^{2}}, 
  \\
\label{gamma-q}
  -\gas &= -1 + \frac{\qs}{\Cs^{2}}\, . 
\end{align}
Alternatively, these consistency conditions can be obtained by inspecting~\eqref{H-iso} for the requirements on $\beta,\gamma$ such that $w$ can be time-independent.

Before continuing, let us pause to develop some physical intuition for eqs.~\eqref{beta-q} and \eqref{gamma-q}. 
Following Ref.~\cite{Mueller:1999pi}, we consider the following phenomenological equation describing the temporal evolution of the average longitudinal momentum for a system undergoing Bjorken expansion 
\begin{align}
\label{pz2-evo}
  \pd_{y}\langle p^{2}_{z}\rangle = - 2 \langle p^{2}_{z}\rangle + 2 D\, .
\end{align}
The average over the phase space weighted by the distribution $\langle\ldots\rangle$ is defined in eq.~\eqref{average}. 
The first and second terms on the right hand side of eq.~\eqref{pz2-evo} account for the effects of the longitudinal expansion and diffusion in momentum space with diffusive constant $D$, respectively. 
In the scaling regime, we further have $\langle p_z^2 \rangle = c_0 \Cs^2$ where $c_{0}$ is a constant of order one. Using the definition of $\g$ in eq.~\eqref{exp-def}, eq.~\eqref{pz2-evo} becomes
\begin{align}
\label{gamma-D}
  -\gas = -1 + \frac{D}{c_{0}\Cs^{2}}\, , 
\end{align}
which is equivalent to eq.~\eqref{gamma-q} with $D\propto q_S$.

The physical interpretation of eq.~\eqref{gamma-q} is now quite clear. 
Recalling $\g$ is the rate of change of the characteristic longitudinal momentum $C$,
eq.~\eqref{gamma-D} indicates that it is given by the combined effects of longitudinal expansion and momentum diffusion. 
Eq.~\eqref{beta-q} can be interpreted similarly in term of transverse momentum diffusion.

We can write down evolution equations for the scaling exponents by differentiating eqs.~\eqref{beta-q} and~\eqref{gamma-q} with respect to $y$:
\bes
 \label{evo}
\begin{align}
   \pd_{y}\betas&= (\dot{q}_{S}+2\betas)\betas\, ,
   \\
   \pd_{y}\gas&= -(\dot{q}_{S}+2\gas)(1-\gas)\, . 
\end{align}
\ees
The evolution of $\als$ is determined from that of $\betas,\gas$ by eq.~\eqref{alpha-relation}.

To close the system of equations~\eqref{evo}, we need to express $\dot{q}_{S}$ in terms of $\betas,\gas$. Substituting the scaling form~\eqref{prescaling-1} for the distribution function into the definition of $q$~\eqref{q-def} yields
\begin{align}
\label{q1}
  \qs= \lambda_0\, \lcb \le(
  c_{a}\tau \As^{2}\Bs^{2}\Cs + \tau n \ri)\, , 
\end{align}
where the time-independent constant $c_{a}$ is given by
\begin{align}
  c_{a}&=\int^{\infty}_{0}\, \frac{d\zeta}{2\pi}\,\zeta \int^{\infty}_{-\infty}\frac{d\xi}{2\pi}\, w^2_{S}(\xi,\zeta)\, .
\end{align}
Using the fact that $\tau n$ is constant from eq.~\eqref{n-evo} along with eq.~\eqref{alpha-relation}, we find
\begin{align}
  \frac{\pd_{y}\le(\tau n + \tau c_{a}\As^{2}\Bs^{2}\Cs\ri)}{\le(\tau n + \tau c_{a}\As^{2}\Bs^{2}\Cs\ri)}
=  \frac{(-1+2\betas+\gas)\tau c_{a}\As^{2}\Bs^{2}\Cs}{\le(\tau_{I} n_{I} + \tau c_{a}\As^{2}\Bs^{2}\Cs\ri)} \, .
\end{align}
As a result, the rate of change of $\qs$ from eq.~\eqref{q1} can be written as
\begin{align}
 \label{q-evo}
  \dot{q}_{S}= \frac{(-1+2\betas+\gas)\tau c_{a}\As^{2}\Bs^{2}\Cs}{\le(\tau_{I} n_{I} + \tau c_{a}\As^{2}\Bs^{2}\Cs\ri)}+\lcbdot\, .
 \end{align}
In sec.~\ref{sec:a-dim}, we shall derive an explicit expression for $\lcbdot$ (see eq.~\eqref{lcb-evo}). 
Since both $\dot{q}_S$ and $\lcbdot$ depend explicitly on $y$ and $\As,\Bs,\Cs$, 
eqs.~\eqref{evo},~\eqref{q-evo}, and ~\eqref{lcb-evo} (together with the relation eq.~\eqref{exp-def}) form a set of closed equations that can be solved for the evolution of the scaling exponents. The solution to these equations (shown in sec.~\ref{sec:exponent-evo}) is the main result of this section.

However, we find it instructive to first consider two limiting cases where the evolution equations~\eqref{evo} are simplified. 
In the first limit, 
we shall assume the distribution is highly-occupied, $\As\gg 1$. 
Since $n\propto \As\Bs^{2}\Cs$ we can neglect the second term in eq.~\eqref{q1} to obtain
\begin{align}
\label{q2}
  \qs\approx \lambda_0\, \lcb\, c_{a} \tau \As^{2}\Bs^{2}\Cs\, . 
\end{align}
Then $\dot{q}_S$ in eq.~\eqref{q-evo} reduces to
\begin{align}
\label{D-evo-1}
  \dot{q}_{S}= -1+2\betas +\gas +\lcbdot\, 
\end{align}
and the evolution equation~\eqref{evo} takes the form
\bes
\label{evo-dense}
\begin{align}
   \textrm{over-occupied ($A_S\gg1$): }\qquad\, \pd_{y}\betas &= \le(\gas+4\betas-1+\lcbdot\ri)\betas \, ,
   \\
   \pd_{y}\gas &= \le(3\gas + 2 \betas -1+\lcbdot\ri)(\gas-1) \, .
 \end{align}
 \ees
In the opposite regime,  we consider a very dilute distribution $\As\ll 1$. 
In this case, the dominant contribution to $\qs$ is from the second term in eq.~\eqref{q1},
\begin{align}
  \qs\approx \lambda_0 \lcb \tau n\, , 
\end{align}
meaning $\dot{q}_{S} = \lcbdot$ since $\tau n$ is time-independent.
Then, we can write eq.~\eqref{evo} as
\bes
\label{evo-dilute}
\begin{align}
   \textrm{dilute ($A_S\ll1$): }\qquad\,\pd_{y}\betas &= \le(2\betas+\lcbdot\ri)\betas \, .
   \\
   \pd_{y}\gas &= (2\gas+\lcbdot)(\gas-1) \, .
 \end{align}
 \ees
These simplified evolution equations~\eqref{evo-dense} and~\eqref{evo-dilute} will be used in the next section to discuss the fixed points of the scaling evolution.

Finally, we emphasize that the evolution equations are derived by assuming the simplified collision integral~\eqref{CIa-1}. 
As we have argued in Appendix~\ref{sec:Ib}, 
this simplification describes well the scaling evolution of hard gluons with $A \geq 1$. 
In this sense, we should be cautious when applying those equations to a dilute system with $A\leq 1$. 
Nevertheless, we notice numerically in sec.~\ref{sec:scaling-num} that scaling exponents extracted using eq.~\eqref{CIa-1} agree well with those from solving the full FP equation even near the dilute fixed point.
We therefore expect that the evolution equations shown here be able to describe scaling in the dilute regime, at least qualitatively.

\subsection{Fixed points
\label{sec:fixed-point}
}

Before solving the evolution equations~\eqref{evo}, let us first identify the possible (non-thermal) fixed points, which correspond to the values of exponents $\betas,\gas$ such that the right hand side of eq.~\eqref{evo} vanishes. 
These fixed points play an important role in characterizing the scaling evolution. 
We will first assume that $\lcbdot =0$, and later in this section illustrate the qualitative implications of a non-zero but constant $\lcbdot$. In subsections~\ref{sec:a-dim} and \ref{sec:exponent-evo} we will derive and then use self-consistent equations for $\lcbdot$.

We begin our discussion by considering perhaps the simplest possibility
\begin{align}
\label{FS-FP}
  &\textrm{Free streaming:}\qquad\, 
  (\als,\betas,\gas)=(0,0,1)\, . 
\end{align}
These exponents automatically make the right hand side of eq.~\eqref{evo} vanish and characterize the free streaming fixed point.
Indeed, 
 \begin{align}
 \label{FS-sol}
    f_{{\rm F.S.}}(\pT,\pz;\tau)= f_{I}(\pT,\left(\frac{\tau}{\tau_{I}} \right)\pz)
\end{align}
solves the Boltzmann equation~\eqref{kin} in the collisionless limit for a generic initial condition
$f(\pT,\pz;\tau=\tau_{I})= f_{I}(\pT,\pz)$.
From the free-streaming solution~\eqref{FS-sol}, 
we can read the corresponding exponents~\eqref{FS-FP} directly.

Next, 
we consider the case with $\gas< 1, \betas =0$.
We note from \eqref{beta-q} that since $\qs$ is finite, 
when we say $\betas=0$ we mean $\qs\ll \Bs^{2}$.
When the typical occupancy is large $\As \gg 1$, 
we can use eq.~\eqref{evo-dense}, which reproduces the BMSS fixed point~\cite{Baier:2000sb} in the absence of $\lcbdot$
\begin{align}
\label{BMSS}
  &\,\textrm{BMSS:}\qquad\, 
  (\als,\betas,\gas)=(-2/3,0,1/3)\, . 
\end{align}
In fact,
for $\gas(y=0)=\g_{I}$ and $\betas=0,\lcbdot=0$, 
we can solve eq.~\eqref{evo-dense} analytically
\begin{align}
\label{g-analytic}
    \gas = \frac{(\g_{I}-1)+e^{-2y}(1-3\g_{I})}{3(\g_{I}-1)+e^{-2y}(1-3\g_{I})}\, . 
\end{align}
For sufficiently large $y$, 
$\gas$ will flow from $\g_{I}$ to the BMSS fixed point value $1/3$. 
The only exception to this would be if $\gas$ starts at the unstable fixed point $\gamma_I = 1$, in which case the solution would stay there forever. Dynamically, however, the original evolution equation for $\gas$~\eqref{C-cond} sets $\gas <1$ always, and therefore the system always flows to the BMSS fixed point in the regime $f \gg 1$.

Finally, we turn to the situation where the system becomes dilute during its expansion, $\As\ll 1$. 
We then read the third fixed point from eq.~\eqref{evo-dilute}: 
\begin{align}
\label{dilute}
&\,\textrm{Dilute:}\qquad \gas=\betas=0\, 
. 
\end{align}
In this limit, the solution to eq.~\eqref{evo-dilute} reads
\begin{align}
    \betas = -\frac{1}{2y - \beta_I^{-1} } \, , & & \gas = \frac{1}{ 1 + \left( \gamma_I^{-1} - 1 \right) e^{2y} }\, ,
\end{align}
which approaches $(\betas, \gas)=(0,0)$ at late times.

The careful reader might ask if imposing the condition $2\betas = -{\dot q}_{S}$ leads to additional fixed points, but it does not, provided $\lcbdot = 0$.
In the limit $\As \gg 1$ with constant $\lcb$, both $-1+\gas$ and $\betas$ are negative by virtue of the consistency conditions~\eqref{beta-q},~\eqref{gamma-q} while ~eq.~\eqref{D-evo-1} implies that $\dot{q}_S < 0$, meaning there is no solution to $2\betas=-\dot{q}_S$. 
In the dilute regime 
$\dot{q}_S\sim 0$, and so $2\betas + \dot{q}_S = 0$ reduces to $\betas = 0$.

From the possible fixed points discussed above, we anticipate three possible scenarios during the far-from-equilibrium stage of the evolution:  
\begin{enumerate}
\item Scenario~I: The expanding plasma evolves from the free streaming fixed point to the BMSS fixed point. 
After that, thermalization occurs. 
This scenario has been discussed extensively in the literature.  

However, because of the presence of the dilute fixed point~\eqref{dilute}, there are two additional possibilities. 
 
\item Scenario~II: Scaling exponents first approach the BMSS fixed point, and then move to the dilute fixed point.

 \item Scenario~III: The exponents are not attracted to the BMSS fixed point, but transit directly to the dilute fixed point. 
\end{enumerate}

These scenarios are consistent with what we observed in the numerical solutions to the FP equation in sec.~\ref{sec:scaling-num}.

The key new finding in this section is the identification of the dilute fixed point~\eqref{dilute}. 
The presence of this new fixed point leads to two additional scenarios in the far-from-equilibrium evolution, namely Scenarios II and III described above. 
To appreciate the underlying physics, we inspect the relation between $\betas,\gas$ and momentum diffusion rate $\qs$~\eqref{beta-q}, \eqref{gamma-q}. 
The vanishing of $\gas$ around this fixed point means the characteristic longitudinal momentum $\Cs$ approaches a constant value, implying that the change of the typical longitudinal momentum due to the expansion is balanced by the momentum diffusion $\qs$. 
On the other hand, the diffusion of transverse momentum is still small compared with its typical value $\Bs$ so that $\betas\to 0$.

For the dilute fixed point to be realized, the typical occupancy number should become small before thermalization. 
Since the occupancy number is characterized by $\As$, 
we estimate the time scale at which the system becomes dilute by $\As(\tau_{{\rm di}}) \sim 1$. Using the relation between $\As$ and $\als$ in eq.~\eqref{exp-def} and estimating $\als \sim -1$ gives
 \begin{equation}
 \label{y-di}
   \tau_{{\rm di}}\sim\tau_{I} A_{I}\, ,
 \end{equation} 
indicating that $\tau_{{\rm di}}$ becomes shorter with smaller occupancy.
 Parametrically, we can take $\tau_I Q_s$ to be of order one and consequently $Q_{s}\tau_{{\rm di}}\sim A_I =\sigma_0/g_s^{2}$.
The thermalization time in the FP equation is parametrically $Q_{s}\tau_{{\rm th}}\sim \exp(1/g_s^2)$~\cite{Bjoraker:2000cf}. 
Comparing the two, we anticipate that there is a range of $g_s$ for which $\tau_{{\rm di}}< \tau_{{\rm th}}$ so that Scenario~II and III would occur. 
This expectation has been confirmed numerically in Fig.~\ref{fig:FP-exp}. 
For $(g_s,\sigma_0)=(0.1,0.6)$ as in the middle panel of Fig.~\ref{fig:FP-exp}, $\tau_\text{di}\sim 60\tau_I$ is in good correspondence to the time scale when the exponents turn toward the dilute fixed point. 
For $(g_s,\sigma_0)=(1/3,0.1)$, as in the right panel of Fig.~\ref{fig:FP-exp}, $\tau_\text{di} \sim \tau_I$ and there is no approach to the BMSS fixed point.
We note, however, that the dominant thermalization processes in the FP equation and QCD EKT are different, and the thermalization time in the latter theory is parametrically shorter. 
Therefore, it would be interesting to examine if Scenario~II and III are relevant for QCD EKT. We leave this as an open question for future investigation.

We now turn to discussing the effect of $\lcbdot$ on the fixed points. 
We shall first discuss the modifications to the BMSS fixed point. 
For this discussion, it is sufficient to set $\betas=0$ and use eq.~\eqref{evo-dense} to find that
\begin{align}
\label{g-lcb}
  \gas = \frac{1}{3}\le(1-\lcbdot\ri)\, ,
\end{align}
which clearly indicates that the scaling exponent $\gas$ differs from the BMSS value $1/3$ due to $\lcbdot$. We interpret the contribution from $\lcbdot$ as an ``anomalous dimension" correction to BMSS scaling exponents, in analogy with the fact that the renormalization group flow in field theories can generate an ``anomalous'' correction to the scaling exponents of correlation functions. 
Remarkably, we will see in the coming section (see eq.~\eqref{delta-gamma}) that this anomalous dimension does depend on the initial values of $A, B, C$, in contrast to the BMSS fixed point exponents which are independent of the initial conditions.\footnote{
According to the general theory of self-similar evolution developed by Barenblatt, 
the situation that scaling exponents are not fully fixed by dimensional analysis but depend on initial conditions is referred to as self-similarity of the second kind~\cite{barenblatt1996scaling}.
The anomalous dimension correction observed in this work fits into this classification.
See also Ref.~\cite{PhysRevLett.64.1361} for an example of the emergence of an anomalous dimension in non-linear diffusive processes.}

Following similar steps, we obtain the effects of $\lcbdot$ on scaling exponents near the dilute fixed point. In this case,
we see from \eqref{evo-dilute} that the presence of $\lcbdot$ also introduces an anomalous dimension correction to the dilute fixed point
\begin{equation}
\label{delta-gamma-dilute}
    \gas = \betas = - \frac12 \lcbdot.
\end{equation}
The fixed point with $(\betas, \gas) = (0, - \lcbdot/2)$ is also possible, but is unstable under time evolution.

\begin{figure}
    \centering
    \includegraphics[width=0.45\textwidth]{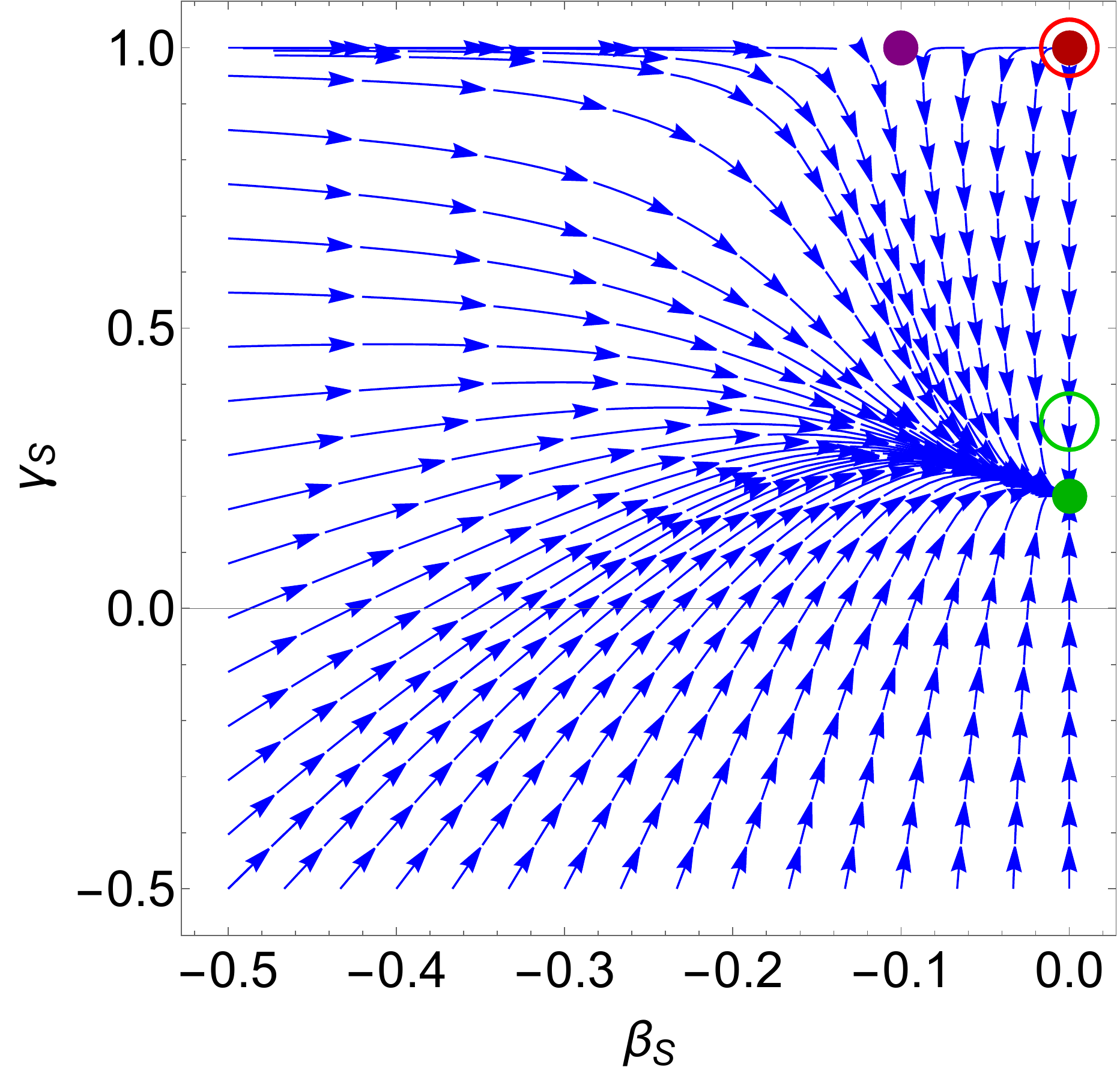} \hspace{0.03\textwidth}
    \includegraphics[width=0.45\textwidth]{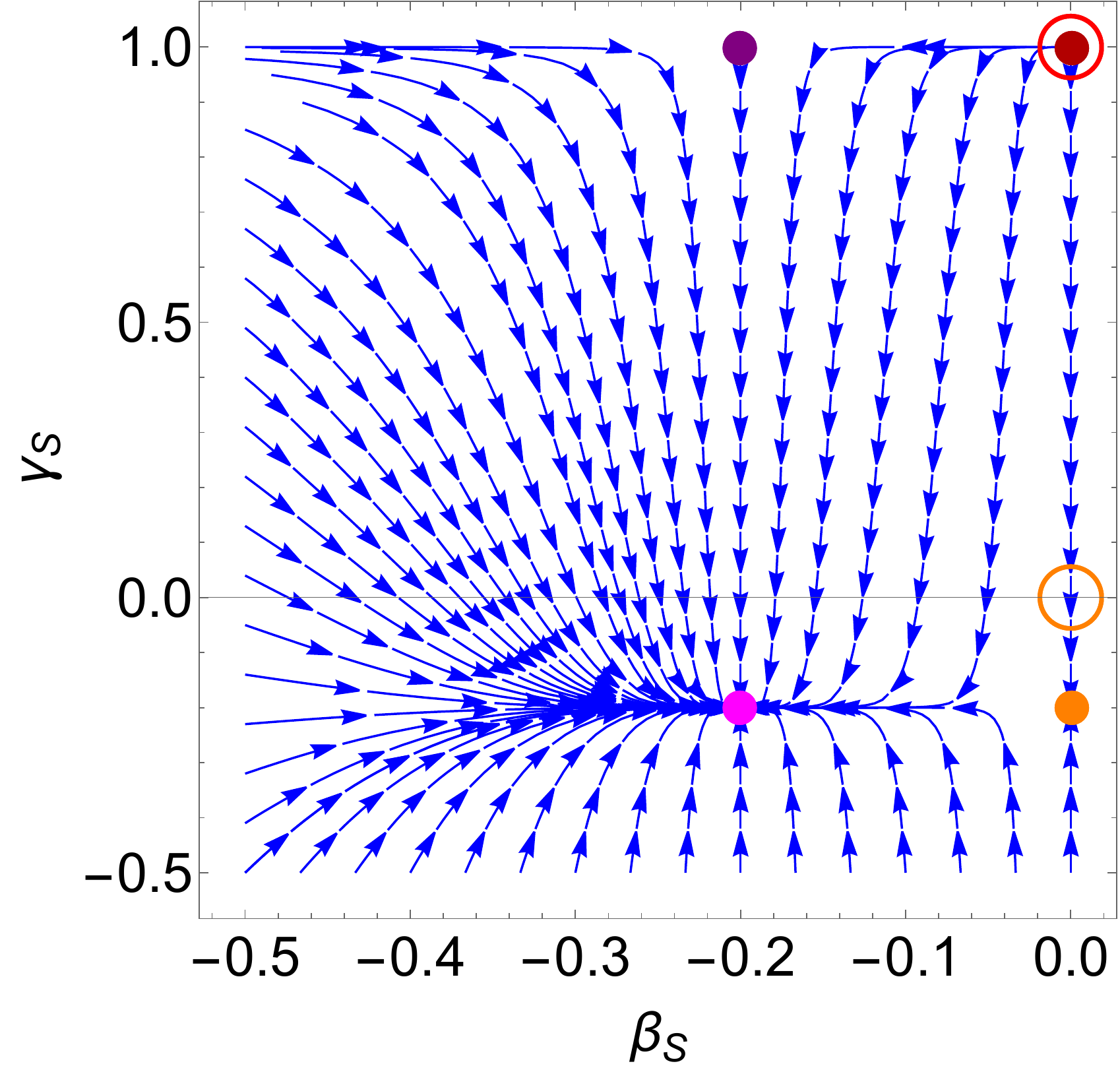}
    \caption{Stream flow of the scaling exponents. Blue arrows represent the flow of the scaling exponents $\betas,\gas$ under time evolution. (left) $f\gg 1$, (right) $f\ll 1$. For illustrative purposes, we set $\lcbdot = 0.4$ and show the corresponding fixed points in filled circles. Fixed points of the evolution equations with $\lcbdot = 0$ are shown as open circles.  Red and purple markers show the free-streaming fixed point with the ``anomalous'' correction and the one without the ``anomalous'' correction, respectively. Green markers show the BMSS fixed point. The orange and pink markers show the dilute fixed point with the anomalous correction in both $\betas$ and $\gas$ and the one with only the ``anomalous'' correction in $\gas$, respectively.
    }
    \label{fig:flow}
\end{figure}

To summarize this section, we show in Fig.~\ref{fig:flow} the fixed points and flow of exponents in the $(\betas, \gas)$ plane. Though $\lcbdot$ is generally time-dependent, for illustrative purposes we take it to be constant, here fixed to $\lcbdot = 0.4$ for visual clarity. For comparison, we show the fixed points with $\lcbdot = 0$ in open circles.
The left panel shows the overoccupied case, where $f \gg 1$. Here, at early times (earlier in the time evolution flow), the free-streaming fixed point with the ``anomalous'' correction is preferred over the ``non-anomalous'' one (which has no $\lcbdot$-dependent corrections), in the sense that flow lines between them go from the ``non-anomalous'' fixed point towards the ``anomalous'' fixed point. At late times, the exponents flow to the BMSS fixed point, which also includes an anomalous correction due to $\lcbdot$ (albeit that this fixed point has no ``non-anomalous'' counterpart). 
On the other hand, the right panel shows the dilute case, with $f \ll 1$. In this situation, at early times, the free-streaming fixed point with the ``anomalous'' correction is again dynamically preferred over the ``non-anomalous'' one. At late times, the exponents flow to the dilute fixed point that includes the anomalous correction due to $\lcbdot$ for both $\beta$ and $\gamma$.

Therefore, we see that the effects of $\lcbdot \neq 0$ are qualitatively relevant to properly understand the exponents near the stable, attractive fixed points. Hence, a more detailed investigation into the consequences of having a nonzero $\lcbdot$ is warranted.

\subsubsection{The Coulomb logarithm}
\label{sec:a-dim}

We will now obtain an explicit expression for $\lcbdot$. We assume the scaling function $w_{S}$ takes the Gaussian form~\eqref{w-sol1} and find
 \begin{align}
 \label{mD-in-ABC}
    m_D^2 &= 4 N_{c} g^2 c_{b}(r_{S}) \As \Bs \Cs\, ,
 \qquad
 \langle\pT^{2}\rangle=2\Bs^{2}
  \, ,
 \end{align}
 where $r_{S}=\frac{\Cs}{\Bs}$, and 
 \begin{align}
 \label{cb-vs-r}
   c_{b}(r_{S})&=
   \int^{\infty}_{0}\, \frac{d\zeta}{2\pi}\,\zeta \int^{\infty}_{-\infty}\,\frac{d\xi}{2\pi} \frac{1}{\sqrt{\zeta^2+r^{2}_{S}\xi^2}}
  w_{S}(\zeta,\xi)\, 
  =  \frac{1}{2\pi^{2}}\, \frac{\arccos(r_{S})}{\sqrt{1-r^{2}_{S}}}\,. 
 \end{align}
Using the definition~\eqref{lcb-0}, we now have
 \begin{align}
 \label{lcb-1}
   \lcb=
   \log \left( \frac{\sqrt{\langle p^{2}_{\perp}\rangle}}{m_{D}} \right)
   =
   \frac{1}{2}\log \le[
   \frac{\Bs}{2 N_{c}g^2_{s} c_{b}(r_{S}) \As\Cs}\, 
   \ri]\, 
 \end{align}
  which in turn gives
 \begin{align}
 \label{lcb-evo}
\lcbdot=\frac{1}{2 \lcb}\le[1-3\betas - \frac{c'_{b}(r_{S})r_{S}}{c_{b}(r_{S})}(\betas-\gas)\ri]\, .
 \end{align}

To obtain a more explicit expression for eq.~\eqref{g-lcb}, 
we use that $\tau \As \Bs^2 \Cs = \tau_I A_I B_I^2 C_I$ is time-independent due to eq.~\eqref{scaling-relation}, and we take $B \approx B_I$ to be approximately constant. This is arbitrarily accurate near the BMSS fixed point, since $\betas=0$ there, and is a reasonable approximation near the dilute fixed point, up to $\lcb$-dependent corrections (because there we have $\betas = -\lcbdot/2$). We get
\begin{align}
  \frac{\Bs}{\As\Cs}=
  \frac{e^{y}}{A_{I}}\, \frac{B_{I}}{C_{I}}\, .
\end{align}
The argument of the $\log$ in eq.~\eqref{lcb-1} now becomes 
\begin{align}
  \frac{\Bs}{2 N_{c}g^2c_{b}(r_{S}) \As\Cs}
  \approx \frac{1}{ 2 N_{c} g^2_{s} c_{b}(r_{S}) }\frac{e^{y}}{A_{I}}\, \frac{B_{I}}{C_{I}}\, ,
\end{align}
so that~eq.~\eqref{lcb-1} gives
\begin{align}
\label{lcb-2}
  \lcb \approx \frac{1}{2}\le[
  y+ \log \le(\frac{B_{I}}{2 N_{c} g^{2}_{s} c_b(r_{S}) A_{I} C_{I}}\ri) 
  \ri]
 \approx
  \frac{1}{2}y + \lcb^I\, ,
\end{align}
where $\lcb^I$ is the value of $\lcb$ at $y=0$,
\begin{align}
\label{lcb-I}
\lcb^I \approx \frac{1}{2}\, \log \le(\frac{ 2\pi   }{g^{2}_{s} N_{c}A_{I}}\, \frac{B_{I}}{C_{I}}\ri)\, . 
\end{align}
We have assumed $r_{S}\to 0$ so that $c_{b}\approx c_{b}(0)=1/(4\pi)$ does not evolve in time. 
Therefore
\begin{equation}
    \lcbdot = \frac{1}{2 \lcb}
\end{equation}
and the correction to the BMSS value now reads
\begin{align}
\label{delta-gamma}
  \gas- \frac{1}{3}=-\frac{1}{3}\lcbdot\approx
  -\frac{1}{3\le(y+2 \lcb^{I}\ri)}\, . 
\end{align}
As noted above, it is remarkable that, unlike the BMSS fixed point exponents, the anomalous dimension in eq.~\eqref{delta-gamma} depends on the initial values of $A, B, C$ through its dependence on $\lcb^{I}$.

\subsection{Solutions 
\label{sec:exponent-evo}
}

In this section, we shall showcase the solutions to eq.~\eqref{evo}, with $\dot{q}_S$ and $\lcbdot$ given by eqs.~\eqref{q-evo} and \eqref{lcb-evo}, respectively.
Our goal is to illustrate the three different scenarios for the temporal behavior of the scaling exponents described in sec.~\ref{sec:fixed-point} and the impact of the time evolution of $\lcb$ on the fixed points.

To solve eq.~\eqref{evo}, we specify initial conditions by matching the scaling form of the distribution eq.~\eqref{f-w} with the initial condition \eqref{fI} for $\xi_{0}=2$ by choosing $A_{I}=\sigma_{0}/g^{2}_{s}, B_{I}=Q_{s}/\sqrt{2}, C_{I}=Q_{s}/(2\sqrt{2})$. 
The initial values of the exponents $\gamma_I, \beta_I$ are fixed by the consistency conditions~\eqref{gamma-q} and~\eqref{beta-q}, with $q_{S}$ evaluated using~\eqref{q1}.
With $\sigma_0$ fixed, the typical occupation number is controlled entirely by the coupling constant $g_s$. 
Therefore, we anticipate that the transition from Scenario~I to Scenario~III through Scenario~II occurs by increasing $g_{s}$.

\begin{figure}[t]
    \centering
        \includegraphics[width=.69\textwidth]{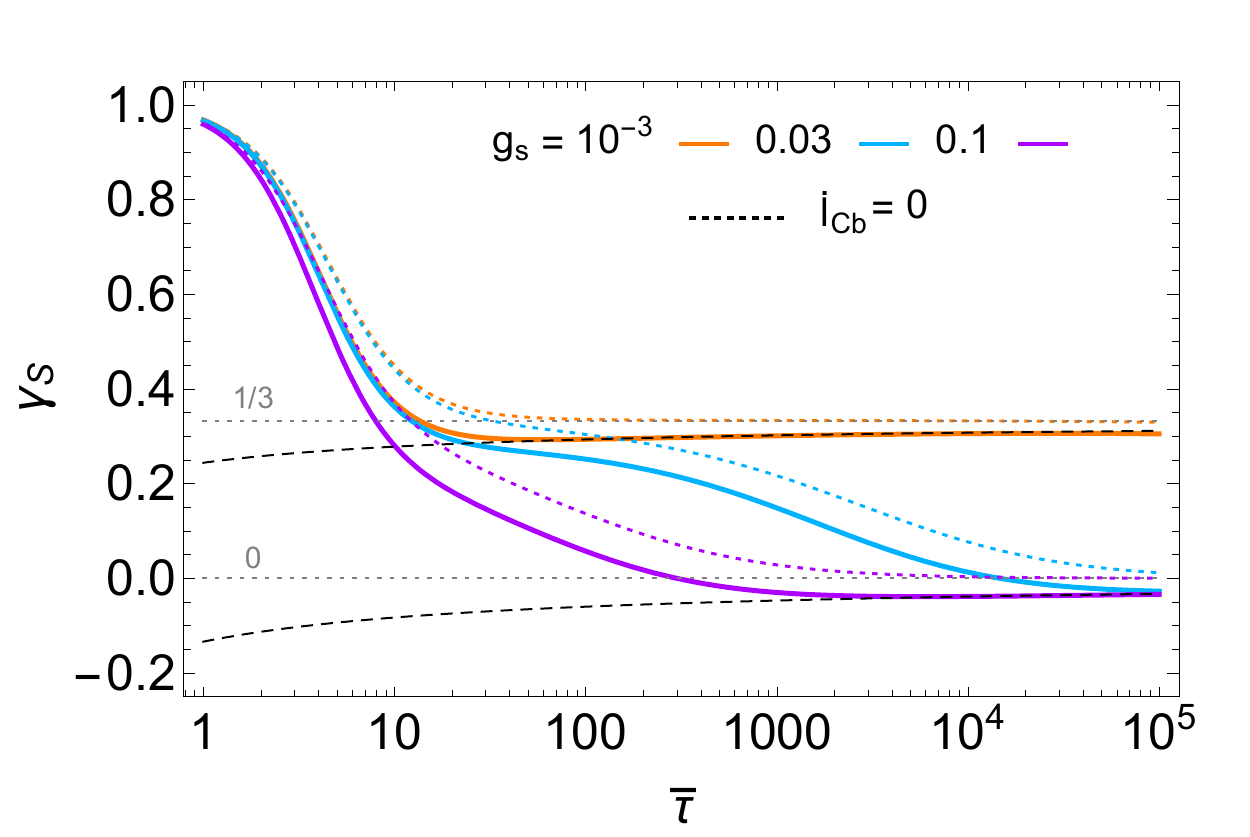}
        \includegraphics[width=.69\textwidth]{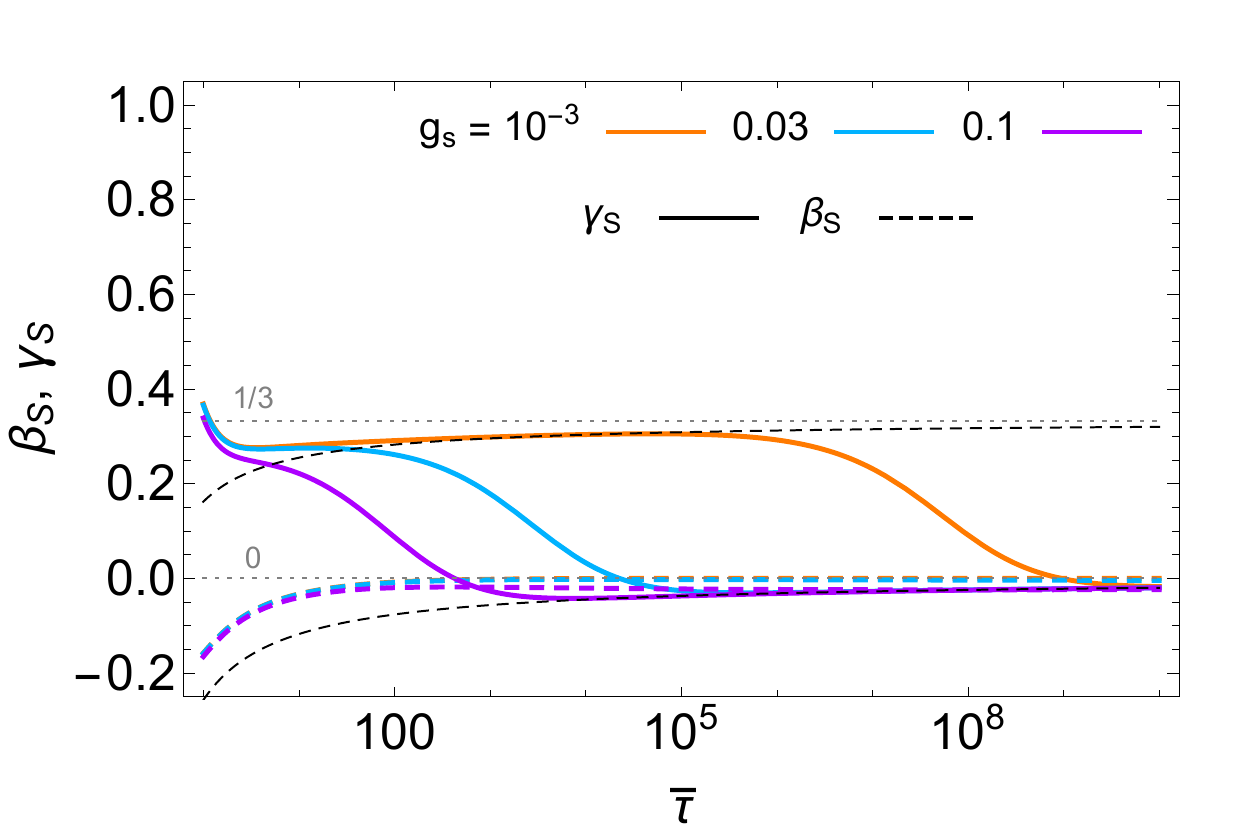}
    \caption{
    \label{fig:scenario}
Evolution of scaling exponents for solutions to eq.~\eqref{evo} for $\gas$ with representative values of the coupling constant $g_s=10^{-3}$ (orange), $0.03$ (blue), and $0.1$ (purple) are shown in solid lines, for $\sigma_0=0.1$ (top) and $\sigma_0=0.6$ (bottom). 
The evolution of $\betas$ is shown by colored dashed lines in the bottom panel ($\betas=0$ in the top panel).
In the top panel, colored dotted lines show solutions with $\lcbdot=0$ for the same set of $g_s$. 
Thin dashed black lines show results for the fixed points including anomalous dimension corrections from eqs.~\eqref{delta-gamma-dilute} and~\eqref{delta-gamma}. 
}
\end{figure}

In Fig.~\ref{fig:scenario}, we show the evolution of the scaling exponents as a function of time for $\sigma_0=0.1$ (left) and $\sigma_0=0.6$ (right), for a range of couplings $g_s$ (indicated by solid colored curves). 
In the left panel we show the evolution of $\gas$ from eq.~\eqref{evo} with $\lcbdot$ given by eq.~\eqref{lcb-evo} and $\sigma_0=0.1$. 
We show only $\gas$ since $|\betas| \lesssim 10^{-3}$ and $\als$ is given by~eq.~\eqref{scaling-relation}. 
For this very small coupling $g_s=10^{-3}$, the scaling exponents approach the BMSS fixed point as in Scenario~I. 
For an intermediate value of the coupling $g_s=0.03$, $\gas$ spends a short time near the BMSS fixed point before transiting to the dilute fixed point as in Scenario~II. 
For larger couplings $g_s=0.1$, $\gas$ goes directly to the dilute fixed point as expected from Scenario~III. 
Therefore we confirm the three scenarios anticipated in the previous qualitative analysis.

It is noteworthy that the late-time values of the exponents at the fixed points are visibly different from the values anticipated in eqs.~\eqref{BMSS} and ~\eqref{dilute}, which are derived by assuming a constant Coulomb logarithm. 
To understand the origin of this deviation, we also show solutions to~eq.~\eqref{evo} with $\lcbdot=0$ in dotted colored curves.
When $\lcbdot=0$ we see that the asymptotic values of the exponents agree exactly with eqs.~\eqref{BMSS} and ~\eqref{dilute}, thus confirming that the deviation arises from the time evolution of $\lcb$.
Indeed, the modification of the asymptotic values of $\gamma_S$ is quantitatively well-described by eq.~\eqref{delta-gamma}, which is shown in thin dashed lines.

%
%
%
%

In the right panel of Fig.~\ref{fig:scenario} we show the evolution of $\betas$ and $\gas$ for $\sigma_0=0.6$. 
The evolution of $\gas$ is again shown in solid colored lines and the evolution of $\betas$ is shown in dashed colored lines. 
In this case, $\betas$ can be a few percent, but this non-zero value of $\betas$ has a small impact on the evolution of $\gas$. 
We note that we show a larger time interval in the right panel than we did in the left. 
On this longer timescale, we see that $g_s=10^{-3}$ eventually transits from the BMSS fixed point to the dilute fixed point, as expected since $\tau_\text{di}\sim (\sigma_{0}/g_s^{2})\tau_{I}\sim 10^{5} \tau_{I}$.
In addition to the modification of the fixed point for $\gas$ discussed in the previous paragraph, we also see that the fixed point for $\betas$ is modified from $0$.
The fixed points for $\gas$ are quantitatively described by eq.~\eqref{delta-gamma} in both the left and right panels of Fig.~\ref{fig:scenario}. For $g_s=0.1$, the late-time fixed point for $\betas$ in the right panel agrees quantitatively with ~eq.~\eqref{delta-gamma-dilute}. 
Since $\betas$ is close to zero, we note that it can take a long time for the fixed point to be reached.
We anticipate that at later times, $\gas=\betas$ would also be realized for the smaller couplings in the right panel of Fig.~\ref{fig:scenario}.

\subsection{Comparison to solutions of kinetic theory
\label{sec:kinetic}
}

\begin{figure}[t]
  \centering
  \includegraphics[width=.69\textwidth]{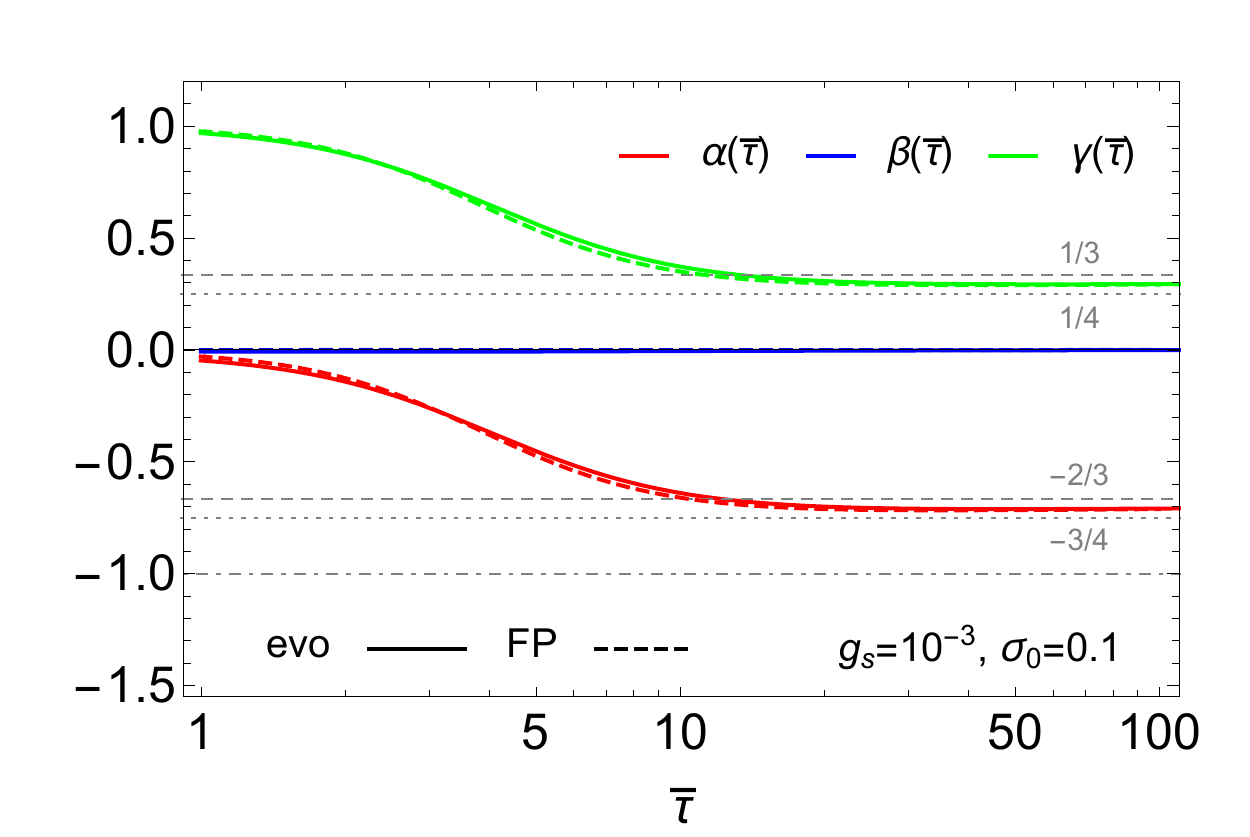}
    \includegraphics[width=.69\textwidth]{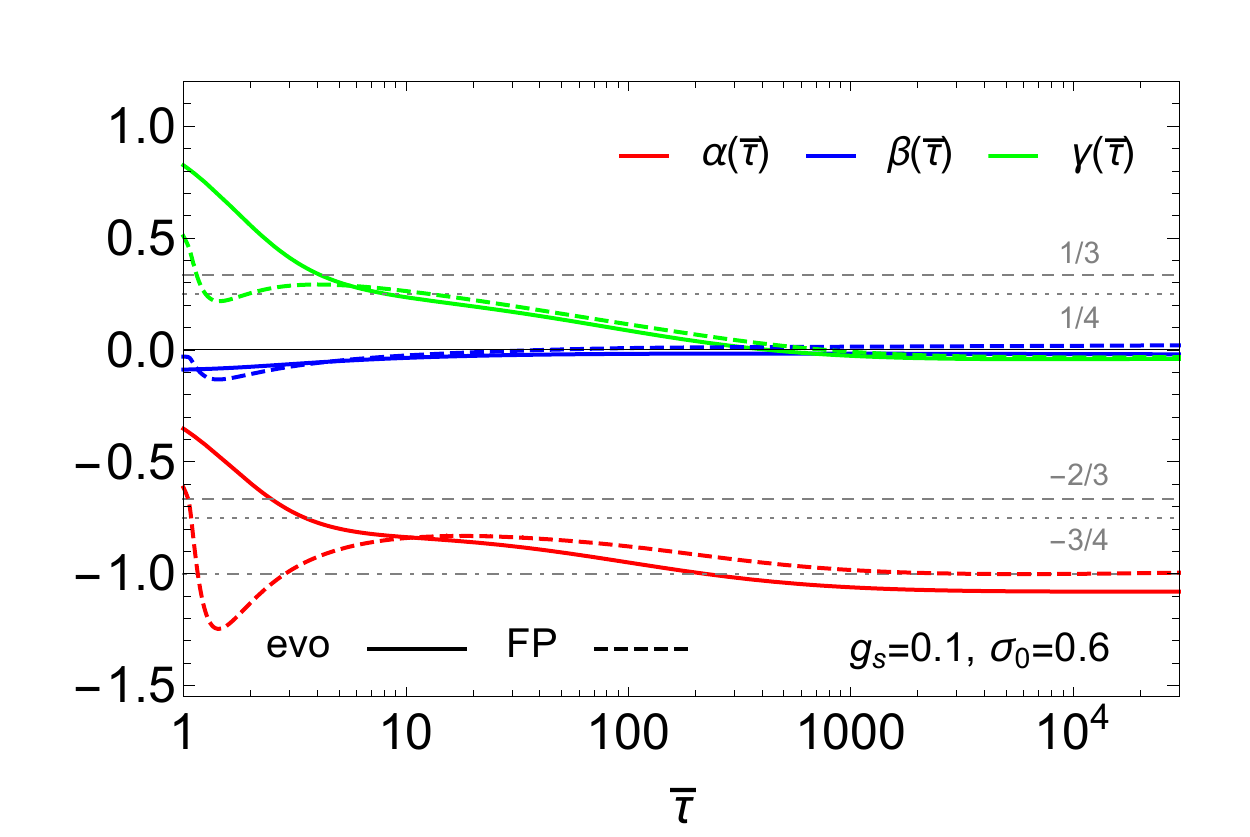}
  \caption{
\label{fig:FP}
We compare the evolution of the scaling exponents from eq.~\eqref{evo} (solid curves) with results from the FP equation (dashed).
In the top panel we take the same initial distribution function for both the evolution equations and the FP equation
at $\tau_I$. In the bottom panel we specify initial conditions for~eq.~\eqref{evo} at $\bar{\tau}_S=3.1$ (see text for details), corresponding to the approximate time for scaling (see the middle panel of Fig.~\ref{fig:FP-exp}). For clarity of presentation, in both panels the dashed curves are the average of exponents computed from different sets of moments of the distribution function.
  }
\end{figure}

Finally, we compare the evolution of scaling exponents obtained from eq.~\eqref{evo} to those extracted from full solutions to kinetic theory.
In Fig.~\ref{fig:FP} we compare to solutions of the FP equations with two different combinations of $\le(\s_{0},g_{s}\ri)=
\le(10^{-3},0.1\ri),\le(0.1,0.6\ri)$. 
These FP results have already been presented in Fig.~\ref{fig:FP-exp} (left) and (middle), and are reproduced in Fig.~\eqref{fig:FP}.
We first note that the solutions to eq.~\eqref{evo} are indistinguishable from the curves for $\I_b=0$ in Fig.~\ref{fig:FP-exp} with the same initial conditions for the distribution function, so these are not shown. 
We emphasize that the evolution equations~\eqref{evo} only apply to the evolution in the scaling regime.

For $(g_s,\sigma_0)=(10^{-3},0.1)$, we see from Fig.~\ref{fig:FP-exp} (left) that the distribution function is approximately scaling from $\tau_I$. In this case we can compute initial conditions for eq.~\eqref{evo} at $\tau_I$ directly from the initial distribution~\eqref{fI}. The results are shown in Fig.~\ref{fig:FP} (left). We observe remarkable agreement between the results from solving eq.~\eqref{evo} and from numerically solving the FP equation. However, for a distribution function that is not initially scaling, in general we should specify initial conditions for eq.~\eqref{evo} after the distribution function has taken the scaling form. This is the case for $(g_s,\sigma_0)=(0.1,0.6)$ where we see substantial deviations from scaling at early times in Fig.~\ref{fig:FP-exp} (middle). We estimate the time to reach the scaling form to be $\tau_{S}/\tau_I \approx 3.1$. 
Then we can estimate $\As,\Bs,\Cs$ from the distribution function at $\tau_{S}$ using $n = \As \Bs^2 \Cs/(2\pi)^{3/2}$, $\langle p_T^2 \rangle = 2 \Bs^2$, and $\langle p_z^2 \rangle = \Cs^2$, and calculate $\gas, \betas$ at $\tau_{S}$ from the consistency conditions~\eqref{beta-q}, \eqref{gamma-q}.
These results are shown in Fig.~\ref{fig:FP} (right) and show good agreement with numerical solutions to the FP equation in the scaling regime.
These results illustrate that in the scaling regime, 
the evolution of the gluon plasma can be reduced to describing the evolution of scaling exponents, in the manner we have done here.

\begin{figure}[t]
  \centering
  \includegraphics[width=.69\textwidth]{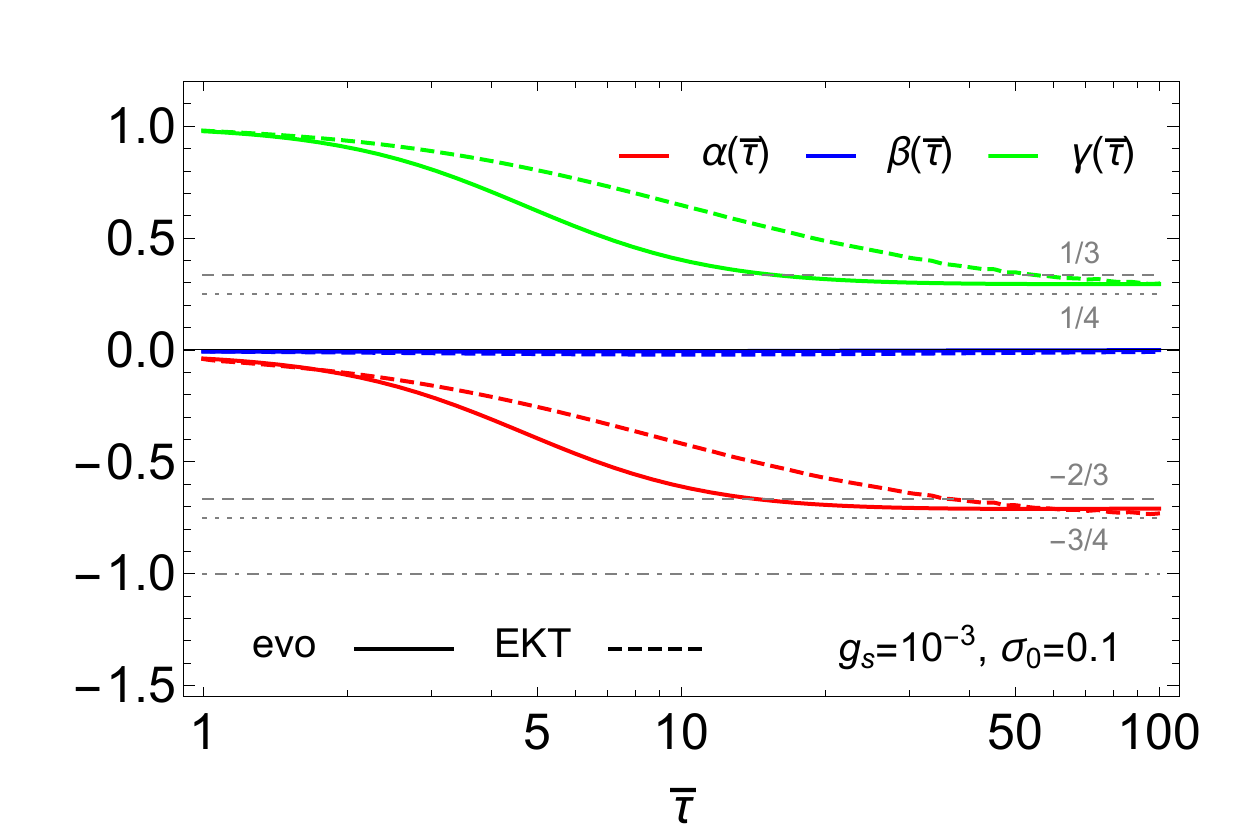}
  \caption{
  \label{fig:EKT}
  The comparison between the evolution of the scaling exponents from eq.~\eqref{evo} (solid curves) with results from QCD effective kinetic theory (EKT) from Ref.~\cite{Mazeliauskas:2018yef} (dashed curves). We take the same initial distribution function as the EKT results at $\tau_I$. For clarity of presentation, EKT results are the average of exponents computed from different sets of moments of the distribution function.
  }
\end{figure}

As we explained earlier,  
we expect that the small-angle scatterings included in the FP equation play the dominant role for the evolution of hard gluons in QCD EKT. We have shown that the collision integral~\eqref{CIa-1} captures the main features of scaling evolution in the FP equation. We therefore wish to compare the evolution equations~\eqref{evo} we have derived based on collision integral~\eqref{CIa-1} to the evolution of scaling exponents in QCD EKT as presented in Ref.~\cite{Mazeliauskas:2018yef}. 
In Fig.~\ref{fig:EKT} we show this comparison for the same initial distribution function~\eqref{fI} and  $(g_s,\sigma_0)=(10^{-3},0.1)$, and observe not only qualitative but also semi-quantitative agreement.

Perhaps the most striking observation that one can draw from Fig.~\ref{fig:EKT} is that the values of $\gas$ and $\betas$ from eq.~\eqref{evo} agree even quantitatively with the exponents from EKT around the BMSS fixed point.
This is highly non-trivial since those asymptotic values are different from their BMSS values.
In Ref.~\cite{Mazeliauskas:2018yef}, 
the authors obtain $\le(\als,\betas,\gas\ri)\approx \le( 0.73, -0.01, 0.29\ri)$ for $(g_s,\sigma_{0})=(10^{-3},0.1)$.\footnote{
In the classical field simulation of Ref.~\cite{Berges:2013eia,Berges:2013fga}, the authors found $\gas= 0.335\pm 0.035$
}
The underlying reason for this deviation from the BMSS value has been the subject of some speculation~\cite{Mazeliauskas:2018yef}.
As we explained in detail in the previous section, the time evolution of $\lcb$ gives rise to an anomalous dimension correction to the scaling exponents (c.f.~eq.~\eqref{delta-gamma}) in the FP equation. 
We therefore propose that the deviation from the BMSS value in QCD EKT may also arise from the time evolution of the ratio between the typical hard scale and typical momentum exchange per collision.

To further test our speculation, we substitute $A_{I}=\sigma_{0}/g^{2}_{s}$ and $B_{I}/C_{I}=2$ into eq.~\eqref{delta-gamma} to estimate the deviation of $\gas$ from BMSS expectation
\begin{align}
\label{delta-gamma-1}
  \delta \gamma \equiv \gas-\frac{1}{3}= - \frac{1}{3 \left(y+ \log \le(\frac{ 4 \pi}{N_{c}\sigma_{0}}\ri) \right)}\, . 
\end{align}
In Ref.~\cite{Mazeliauskas:2018yef}, the evolution of kinetic theory starts at $Q_{s}\tau_I=70$ and ends at $Q_{s}\tau=7000$, meaning we should replace $y$ in eq.~\eqref{delta-gamma-1} with $\log(100)\approx 4.6$. 
For $\sigma_{0}=0.1$ , 
we obtain the correction from $\lcb$ to $\gas$ at the BMSS fixed point: 
\begin{align}
\delta\g = -0.040 \, ,
\qquad
\text{or}\,\,\,\,\,\,\, \gas \approx 0.29
\end{align}
which is in remarkable agreement with the asymptotic value of $\gas$ in EKT.

\section{Summary and outlook
\label{sec:conclusion}
}

In this work we have studied scaling in a Bjorken expanding gluon plasma described by the Boltzmann transport equation under the small-angle approximation, which takes the form of a  Fokker-Planck (FP) equation. 
For hard gluons, we showed that the FP equation features time-dependent scaling behavior that is qualitatively similar to that observed by solving QCD EKT~\cite{Mazeliauskas:2018yef}. 

We then showed that scaling can be interpreted as arising from adiabatic evolution.
With the simplified collision integral~\eqref{CIa-1}, the kinetic equation can be recast into the form of a Schr\"odinger-like equation. 
Adiabaticity, understood as the property that the eigenstates of the corresponding Hamiltonian do not transition into each other, may be attained to a lesser or greater degree depending on the choice of frame. For the particular case we study here, we find that one can choose a rescaling of the momentum coordinates (frame) such that there are no transitions between eigenstates of the corresponding Hamiltonian. 
This means that after some transient time, the excited states have decayed and the distribution function follows the evolution of the instantaneous ground state. 

It is only in this frame that the scaling distribution we observed in the numerical solutions is the ground state.
Without identifying the adiabatic frame, one can still observe scaling phenomena, but the adiabatic nature of the scaling evolution would be obscured. In this sense, we have generalized the notion of the abiabaticity with respect to a fixed set of coordinates $(\tau;p_z,p_\perp)$ to the situation where there exists a ``frame" $(\tau;p_z/C(\tau), p_\perp/B(\tau))$ in which the transition rate from the instantaneous ground state to excited states is suppressed (in this case, zero). We believe that this generalization of adiabaticity may find applications in a broader context.\footnote{
For example, 
consider a time-dependent Hamiltonian in quantum mechanics, and suppose there exists a unitary transformation under which the transformed Hamiltonian evolves slowly. In that case, we can still say that the system described by the original Hamiltonian evolves adiabatically even though this Hamiltonian may change rapidly in time. See Ref.~\cite{scaling-QM} for a similar discussion.
}

From the condition for adiabaticity, we further derived evolution equations for the time dependence of the scaling exponents. 
Our equations can be used to estimate the evolution of typical occupancy and momentum of far-from-equilibrium QGP during the early stages of heavy-ion collisions. In addition to the well-known free-streaming and BMSS fixed points, we found a new ``dilute'' fixed point~\eqref{dilute} that occurs when the typical gluon occupation number becomes small before thermalization.
We also find that the fixed point scaling exponents receive ``anomalous dimension" corrections, arising from the temporal evolution of the Coulomb logarithm, which is determined by the ratio of the hard and soft momentum scales. We compared our results with QCD EKT simulations from Ref.~\cite{Mazeliauskas:2018yef}, and found a striking quantitative agreement on the correction to the BMSS exponent in the two theories. In our analysis, this is precisely due to the time evolution of the Coulomb logarithm. In the view that the FP equation we solve here gives an effective description of time-dependent scaling, in qualitative and semi-quantitative agreement with more sophisticated first-principles QCD EKT simulations, our findings suggest that understanding time evolution in terms of an adiabatic evolution may be a valuable approach for describing far-from-equilibrium QCD plasmas.

The relation between adiabaticity and the non-equilibrium attractor had previously been tested in the simpler single relaxation time approximation~\cite{Brewer:2019oha}. 
Together with the results of Ref.~\cite{Brewer:2019oha}, our finding that the non-thermal scaling evolution of a far-from-equilibrium gluon plasma can be characterized by adiabatic evolution gives compelling support for the claim that the reduction of relevant degrees of freedom in a class of expanding QCD plasmas is due to adiabaticity. Since we have here shown that this class is larger than it was previously known, we anticipate that a similar study of more general kinetic equations will reveal more connections to adiabaticity. More general collision kernels as well as more realistic heavy-ion collision scenarios including radial expansion in the kinetic description of the plasma should be fertile ground for such an investigation.

We hope some of our qualitative lessons, such as the relation between adiabaticity and scaling, and the emergence of anomalous dimension corrections to scaling exponents, might be instructive when studying other dynamical problems. Examples could include the evolution near a critical point based on the Kibble-Zurek framework~\cite{Kibble_1976,Zurek:1985qw,Zurek:1996sj,KZ2012,Mukherjee:2016kyu,Akamatsu:2018vjr}, and turbulent cascades~\cite{frisch1995turbulence} driven by quantum anomalies~\cite{Hirono:2015rla,Mace:2019cqo}. We defer the investigation of these interesting topics to future work.

We note that an analysis of time-dependent scaling exponents in 
Fokker-Planck kinetic theory was performed independently by Aleksandr Mikheev, Aleksas Mazeliauskas, and J\"urgen Berges and made public 
simultaneously to the present manuscript.

\acknowledgments
We thank J\"urgen Berges, Aleksas Mazeliauskas, Aleksandr Mikheev, Krishna Rajagopal, Raju Venugopalan, Urs Wiedemann, and Li Yan for valuable discussions and comments, and Aleksas Mazeliauskas for providing numerical results for the prescaling exponents in EKT. BSH is supported by the U.S. Department of Energy, Office of Science, Office of Nuclear Physics under grant Contract Number DE-SC0011090 (Nuclear Theory research).
YY is supported by the Strategic Priority Research Program of Chinese Academy of Sciences, Grant No. XDB34000000.

\appendix

\section {Numerical implementation}
\label{sec:numerics}

Here we discuss the numerical procedure to solve the FP equation. 
We follow Ref.~\cite{Tanji:2017suk} and write  the FP equation \eqref{kin} in terms of variables $p = \sqrt{p_T^2 + p_z^2}$ and $\kappa = p_z/p$. 
We shall use
\bes
\begin{eqnarray}
&\,	\nabla_{\bf p}^2 f = \frac{\partial^2 f}{\partial p^2} + \frac{1-\kappa^2}{p^2} \frac{\partial^2 f}{\partial \kappa^2} + \frac{2}{p} \frac{\partial f}{\partial p}  - \frac{2 \kappa}{p^2} \frac{\partial f}{\partial \kappa} \, ,
\\
&\,	\nabla_{\bf p} \cdot \left( \frac{{\bf p}}{p} (1+f) f \right) = \frac{1}{p^2} \frac{\partial}{\partial p} p^2 f (1+f)\, ,
\\
&\,
\frac{\partial f}{\partial p_z} = \kappa \frac{\partial f}{\partial p} + \frac{1-\kappa^2}{p} \frac{\partial f}{\partial \kappa}\, .
\end{eqnarray}
\ees
To ease the numerical implementation, we additionally consider $l_p \equiv \log p$ and evolve the quantity $\log f$. 
The FP equation then becomes
\begin{align*}
	& \frac{\partial \log f}{\partial \tau} + \frac{\kappa}{\tau} \left[(\kappa^2-1) \frac{\partial \log f}{\partial \kappa} - \kappa \frac{\partial \log f}{\partial l_p} \right]\\
	&= \lambda_0 \lcb[f] \left( e^{-l_p} \mathcal{I}_b[f] \left[ \frac{\partial \log f}{\partial l_p} + 2 + 2 e^{\log f} \left(1+ \frac{\partial \log f}{\partial l_p} \right) \right] \right.\\
	&\left.+ e^{-2 l_p} \mathcal{I}_a[f] \left[ -2\kappa \frac{\partial \log f}{\partial \kappa} - (\kappa^2-1)\left( \frac{\partial^2 \log f}{\partial \kappa^2} + \left( \frac{\partial \log f}{\partial \kappa}\right)^2\right) + \frac{\partial \log f}{\partial l_p} + \frac{\partial^2 \log f}{\partial l_p^2} + \left( \frac{\partial \log f}{\partial l_p}\right)^2 \right] \right)
\end{align*}
$\mathcal{I}_a$, $\mathcal{I}_b$, and $\lcb$ are integrals that depend on $f$, with $\mathcal{I}_a$ and $\mathcal{I}_b$ defined through eq.~\eqref{eq:IaIb} and $\lcb$ through eq.~\eqref{lcb-0}.
In these coordinates, we note that $\kappa = \cos \theta = p_z/p$ and $\sin \theta = p_T/p$, which give $p_z = e^{l_p} \kappa$ and $p_T = e^{l_p} \sqrt{1-\kappa^2}$. The spherical volume element is $d^3 p = 2 \pi p^2 dp d\kappa = 2 \pi e^{3 l_p} d l_p d\kappa $. The moments \eqref{moment-def} can therefore be written
\begin{equation}
    n_{m,n}(\tau) = \frac{1}{(2\pi)^2} \int d l_p \, d \kappa \, e^{(3+m+n) l_p} (1-\kappa^2)^{m/2} |\kappa|^n f(p_\perp,p_z,\tau)\, .
\end{equation}
The initial condition for the distribution function \eqref{fI} in these coordinates is
\begin{equation}
	\log f(\pT,\pz;\tau=\tau_I) = \log \frac{\sigma_0}{g_s^2} - \frac{e^{2 l_p}(1-(1-\xi_0^2)\kappa^2)}{Q_s^2}\, .
\end{equation}
We use the finite element method in Mathematica's NDSolve to solve the resulting equations in the range $p \in [5 \cdot 10^{-3},4]$, $\kappa \in [0,1]$ (assuming inversion symmetry in $p_z$) and a maximum cell size of $10^{-3}$.

\section{The simplification of collision integral
\label{sec:Ib}
}

In this Appendix, we discuss the simplification of the collision integral~\eqref{eq:small-angle-kernel} in the situation that the ratio of the typical longitudinal momentum to that of transverse momentum, $r=C/B$, is small. 
In addition, we shall justify dropping $\I_{b}$ term for hard gluons under the condition that typical occupancy is large, i.e. $A\gg 1$.

We begin by substituting eq.~\eqref{f-w} into the FP equation~\eqref{kin} with eq.~\eqref{eq:small-angle-kernel} , with which we obtain the equation for the rescaled distribution function $w$ explicitly as
\begin{align}
\label{w-eq-full-1}
    \pd_{y}w=&-\a w + (1-\g)\xi\,w_{\xi}-\beta \zeta w_{\zeta}
    \no 
    \\
    &+\frac{q}{C^{2}}\, 
    \le[
    \le(
    w_{\xi\xi}+r^{2}(\frac{1}{\zeta}w_{\zeta}+w_{\zeta\zeta})
    \ri)+ 
    \frac{2 c_{b} r^{2}}{(c_{a}\, A+d_{a})}\,\frac{1}{\sqrt{\zeta^{2}+r^{2}\xi^{2}}}\le(2+ \xi\pd_{\xi}+\zeta\pd_{\zeta}\ri)\,\le( w+A\, w^{2}\ri)
    \ri]\, ,
\end{align}
where we have used the relation
\begin{align}
    \frac{I_{b}}{I_{a}}=\frac{2 ABC c_{b}}{\le( c_{a}\, A+d_{a}\ri)\, A B^{2} C}=\frac{2 c_{b}}{B (c_{a}\, A+d_{a})}\, .
\end{align}
Here, we have defined
\begin{align}
  c_{a}&\equiv \,\int_{\xi,\zeta}\, w^{2}\, ,
  \qquad
  d_{a}\equiv\int_{\xi,\zeta}\, w\, ,
  \qquad
  c_{b}\equiv 
   \int_{\xi,\zeta} \frac{w}{\sqrt{\zeta^{2}+r^{2}\xi^{2}}} \,, 
\end{align}
and have introduced short-hand notation for the integration over scaling variables
\begin{align}
    \int_{\xi,\zeta}\equiv \int^{\infty}_{-\infty}\, \frac{d\xi}{2\pi}\int^{\infty}_{0}\, \frac{d\zeta}{2\pi}\, \zeta . 
\end{align}
We shall assume $c_{a},d_{a},c_{b}$ to be order one.

Now, we consider eq.~\eqref{w-eq-full-1} in the small $r$ limit.
By looking at eq.~\eqref{q-general}, we can count $\beta$ to be of the order $r^{2}$. 
Therefore at leading order in the small $r$ expansion, we obtain eq.~\eqref{H}, which is equivalent to using the collision integral~\eqref{CIa-0}. 
The correction due to finite $r$ corresponds to terms proportional to $r^{2}$ in eq.~\eqref{w-eq-full-1}.

Next, we consider the limit $A\gg 1$. 
In the regime where $A w \gg 1$ is satisfied, 
eq.~\eqref{w-eq-full-1} is reduced to 
\begin{align}
\label{w-eq-2}
\pd_{y}w&=-\a w + (1-\g)w_{\xi}+\frac{q}{C^{2}}w_{\xi\xi}
    \no 
    \\
    &-\beta \zeta w_{\zeta}+\frac{q}{B^2}
    \le[
    (\frac{1}{\zeta}w_{\zeta}+w_{\zeta\zeta})
    + 
    \,\frac{2c_{b}}{c_{a}}\,\frac{1}{\zeta}\le(2+ \xi\pd_{\xi}+\zeta\pd_{\zeta}\ri)\,w^{2}
    \ri]\, . 
\end{align}
For the tail of the distribution, $\zeta,\xi \gg 1$, we have $w\ll 1$, and then
the last term in the bracket of \eqref{w-eq-2} is small compared with the first term in the bracket and can be dropped. 
This corresponds to setting $\I_{b}=0$, i.e., to using the collision kernel given by eq.~\eqref{CIa-1}.

\bibliographystyle{JHEP.bst}
\bibliography{ref.bib}

\providecommand{\href}[2]{#2}\begingroup\raggedright\begin{thebibliography}{10}

\bibitem{Mazeliauskas:2018yef}
A.~Mazeliauskas and J.~Berges, \emph{{Prescaling and far-from-equilibrium
  hydrodynamics in the quark-gluon plasma}},
  \href{https://doi.org/10.1103/PhysRevLett.122.122301}{\emph{Phys. Rev. Lett.}
  {\bfseries 122} (2019) 122301}
  [\href{https://arxiv.org/abs/1810.10554}{{\ttfamily 1810.10554}}].

\bibitem{akiba2015hot}
Y.~Akiba, A.~Angerami, H.~Caines, A.~Frawley, U.~Heinz, B.~Jacak et~al.,
  \emph{The hot qcd white paper: Exploring the phases of qcd at rhic and the
  lhc},  2015.

\bibitem{Romatschke:2017ejr}
P.~Romatschke and U.~Romatschke, \emph{{Relativistic Fluid Dynamics In and Out
  of Equilibrium}}, Cambridge Monographs on Mathematical Physics, Cambridge
  University Press (5, 2019),
  \href{https://doi.org/10.1017/9781108651998}{10.1017/9781108651998},
  [\href{https://arxiv.org/abs/1712.05815}{{\ttfamily 1712.05815}}].

\bibitem{Busza_2018}
W.~Busza, K.~Rajagopal and W.~van~der Schee, \emph{Heavy ion collisions: The
  big picture and the big questions},
  \href{https://doi.org/10.1146/annurev-nucl-101917-020852}{\emph{Annual Review
  of Nuclear and Particle Science} {\bfseries 68} (2018) 339–376}.

\bibitem{Schenke:2021mxx}
B.~Schenke, \emph{{The smallest fluid on Earth}},
  \href{https://doi.org/10.1088/1361-6633/ac14c9}{\emph{Rept. Prog. Phys.}
  {\bfseries 84} (2021) 082301}
  [\href{https://arxiv.org/abs/2102.11189}{{\ttfamily 2102.11189}}].

\bibitem{Schlichting:2019abc}
S.~Schlichting and D.~Teaney, \emph{{The First fm/c of Heavy-Ion Collisions}},
  \href{https://doi.org/10.1146/annurev-nucl-101918-023825}{\emph{Ann. Rev.
  Nucl. Part. Sci.} {\bfseries 69} (2019) 447}
  [\href{https://arxiv.org/abs/1908.02113}{{\ttfamily 1908.02113}}].

\bibitem{Berges:2020fwq}
J.~Berges, M.P.~Heller, A.~Mazeliauskas and R.~Venugopalan, \emph{{QCD
  thermalization: Ab initio approaches and interdisciplinary connections}},
  \href{https://doi.org/10.1103/RevModPhys.93.035003}{\emph{Rev. Mod. Phys.}
  {\bfseries 93} (2021) 035003}
  [\href{https://arxiv.org/abs/2005.12299}{{\ttfamily 2005.12299}}].

\bibitem{Kurkela:2018wud}
A.~Kurkela, A.~Mazeliauskas, J.-F.~Paquet, S.~Schlichting and D.~Teaney,
  \emph{{Matching the Nonequilibrium Initial Stage of Heavy Ion Collisions to
  Hydrodynamics with QCD Kinetic Theory}},
  \href{https://doi.org/10.1103/PhysRevLett.122.122302}{\emph{Phys. Rev. Lett.}
  {\bfseries 122} (2019) 122302}
  [\href{https://arxiv.org/abs/1805.01604}{{\ttfamily 1805.01604}}].

\bibitem{Schenke:2019pmk}
B.~Schenke, C.~Shen and P.~Tribedy, \emph{{Hybrid Color Glass Condensate and
  hydrodynamic description of the Relativistic Heavy Ion Collider small system
  scan}}, \href{https://doi.org/10.1016/j.physletb.2020.135322}{\emph{Phys.
  Lett. B} {\bfseries 803} (2020) 135322}
  [\href{https://arxiv.org/abs/1908.06212}{{\ttfamily 1908.06212}}].

\bibitem{Giacalone:2020byk}
G.~Giacalone, B.~Schenke and C.~Shen, \emph{{Observable signatures of initial
  state momentum anisotropies in nuclear collisions}},
  \href{https://doi.org/10.1103/PhysRevLett.125.192301}{\emph{Phys. Rev. Lett.}
  {\bfseries 125} (2020) 192301}
  [\href{https://arxiv.org/abs/2006.15721}{{\ttfamily 2006.15721}}].

\bibitem{Kurkela:2021ctp}
A.~Kurkela, A.~Mazeliauskas and R.~T\"ornkvist, \emph{{Collective flow in
  single-hit QCD kinetic theory}},
  \href{https://doi.org/10.1007/JHEP11(2021)216}{\emph{JHEP} {\bfseries 11}
  (2021) 216} [\href{https://arxiv.org/abs/2104.08179}{{\ttfamily
  2104.08179}}].

\bibitem{Arnold:2002zm}
P.B.~Arnold, G.D.~Moore and L.G.~Yaffe, \emph{{Effective kinetic theory for
  high temperature gauge theories}},
  \href{https://doi.org/10.1088/1126-6708/2003/01/030}{\emph{JHEP} {\bfseries
  01} (2003) 030} [\href{https://arxiv.org/abs/hep-ph/0209353}{{\ttfamily
  hep-ph/0209353}}].

\bibitem{Kurkela:2015qoa}
A.~Kurkela and Y.~Zhu, \emph{{Isotropization and hydrodynamization in weakly
  coupled heavy-ion collisions}},
  \href{https://doi.org/10.1103/PhysRevLett.115.182301}{\emph{Phys. Rev. Lett.}
  {\bfseries 115} (2015) 182301}
  [\href{https://arxiv.org/abs/1506.06647}{{\ttfamily 1506.06647}}].

\bibitem{Kurkela:2018xxd}
A.~Kurkela and A.~Mazeliauskas, \emph{{Chemical Equilibration in Hadronic
  Collisions}},
  \href{https://doi.org/10.1103/PhysRevLett.122.142301}{\emph{Phys. Rev. Lett.}
  {\bfseries 122} (2019) 142301}
  [\href{https://arxiv.org/abs/1811.03040}{{\ttfamily 1811.03040}}].

\bibitem{Almaalol:2020rnu}
D.~Almaalol, A.~Kurkela and M.~Strickland, \emph{{Nonequilibrium Attractor in
  High-Temperature QCD Plasmas}},
  \href{https://doi.org/10.1103/PhysRevLett.125.122302}{\emph{Phys. Rev. Lett.}
  {\bfseries 125} (2020) 122302}
  [\href{https://arxiv.org/abs/2004.05195}{{\ttfamily 2004.05195}}].

\bibitem{Du:2020zqg}
X.~Du and S.~Schlichting, \emph{{Equilibration of the Quark-Gluon Plasma at
  Finite Net-Baryon Density in QCD Kinetic Theory}},
  \href{https://doi.org/10.1103/PhysRevLett.127.122301}{\emph{Phys. Rev. Lett.}
  {\bfseries 127} (2021) 122301}
  [\href{https://arxiv.org/abs/2012.09068}{{\ttfamily 2012.09068}}].

\bibitem{Berges:2013eia}
J.~Berges, K.~Boguslavski, S.~Schlichting and R.~Venugopalan, \emph{{Turbulent
  thermalization process in heavy-ion collisions at ultrarelativistic
  energies}}, \href{https://doi.org/10.1103/PhysRevD.89.074011}{\emph{Phys.
  Rev. D} {\bfseries 89} (2014) 074011}
  [\href{https://arxiv.org/abs/1303.5650}{{\ttfamily 1303.5650}}].

\bibitem{Berges:2013fga}
J.~Berges, K.~Boguslavski, S.~Schlichting and R.~Venugopalan, \emph{{Universal
  attractor in a highly occupied non-Abelian plasma}},
  \href{https://doi.org/10.1103/PhysRevD.89.114007}{\emph{Phys. Rev. D}
  {\bfseries 89} (2014) 114007}
  [\href{https://arxiv.org/abs/1311.3005}{{\ttfamily 1311.3005}}].

\bibitem{Tanji:2017suk}
N.~Tanji and R.~Venugopalan, \emph{{Effective kinetic description of the
  expanding overoccupied Glasma}},
  \href{https://doi.org/10.1103/PhysRevD.95.094009}{\emph{Phys. Rev. D}
  {\bfseries 95} (2017) 094009}
  [\href{https://arxiv.org/abs/1703.01372}{{\ttfamily 1703.01372}}].

\bibitem{Schmied_2019}
C.-M.~Schmied, A.N.~Mikheev and T.~Gasenzer, \emph{Prescaling in a
  far-from-equilibrium bose gas},
  \href{https://doi.org/10.1103/physrevlett.122.170404}{\emph{Physical Review
  Letters} {\bfseries 122} (2019) }.

\bibitem{Brewer:2019oha}
J.~Brewer, L.~Yan and Y.~Yin, \emph{{Adiabatic hydrodynamization in
  rapidly-expanding quark-gluon plasma}},
  \href{https://arxiv.org/abs/1910.00021}{{\ttfamily 1910.00021}}.

\bibitem{Gelis:2010nm}
F.~Gelis, E.~Iancu, J.~Jalilian-Marian and R.~Venugopalan, \emph{{The Color
  Glass Condensate}},
  \href{https://doi.org/10.1146/annurev.nucl.010909.083629}{\emph{Ann. Rev.
  Nucl. Part. Sci.} {\bfseries 60} (2010) 463}
  [\href{https://arxiv.org/abs/1002.0333}{{\ttfamily 1002.0333}}].

\bibitem{2001nlin.....11055A}
D.G.~{Aronson}, S.I.~{Betelu} and I.G.~{Kevrekidis}, \emph{{Going with the
  Flow: a Lagrangian approach to self-similar dynamics and its consequences}},
  {\emph{arXiv e-prints} (2001) nlin/0111055}
  [\href{https://arxiv.org/abs/nlin/0111055}{{\ttfamily nlin/0111055}}].

\bibitem{kevrekidis2017revisiting}
P.G.~Kevrekidis, M.O.~Williams, D.~Mantzavinos, E.G.~Charalampidis, M.~Choi and
  I.G.~Kevrekidis, \emph{Revisiting diffusion: Self-similar solutions and the
  $t^{-1/2}$ decay in initial and initial-boundary value problems},  2017.

\bibitem{Baier:2000sb}
R.~Baier, A.H.~Mueller, D.~Schiff and D.T.~Son, \emph{{'Bottom up'
  thermalization in heavy ion collisions}},
  \href{https://doi.org/10.1016/S0370-2693(01)00191-5}{\emph{Phys. Lett. B}
  {\bfseries 502} (2001) 51}
  [\href{https://arxiv.org/abs/hep-ph/0009237}{{\ttfamily hep-ph/0009237}}].

\bibitem{cha95}
P.M.~Chaikin and T.C.~Lubensky, \emph{Principles of Condensed Matter Physics},
  Cambridge University Press, Cambridge (1995).

\bibitem{Mueller:2002gd}
A.H.~Mueller and D.T.~Son, \emph{{On the Equivalence between the Boltzmann
  equation and classical field theory at large occupation numbers}},
  \href{https://doi.org/10.1016/j.physletb.2003.12.047}{\emph{Phys. Lett. B}
  {\bfseries 582} (2004) 279}
  [\href{https://arxiv.org/abs/hep-ph/0212198}{{\ttfamily hep-ph/0212198}}].

\bibitem{Mueller:1999pi}
A.H.~Mueller, \emph{{The Boltzmann equation for gluons at early times after a
  heavy ion collision}},
  \href{https://doi.org/10.1016/S0370-2693(00)00084-8}{\emph{Phys. Lett. B}
  {\bfseries 475} (2000) 220}
  [\href{https://arxiv.org/abs/hep-ph/9909388}{{\ttfamily hep-ph/9909388}}].

\bibitem{Blaizot:2013lga}
J.-P.~Blaizot, J.~Liao and L.~McLerran, \emph{{Gluon Transport Equation in the
  Small Angle Approximation and the Onset of Bose-Einstein Condensation}},
  \href{https://doi.org/10.1016/j.nuclphysa.2013.10.010}{\emph{Nucl. Phys. A}
  {\bfseries 920} (2013) 58} [\href{https://arxiv.org/abs/1305.2119}{{\ttfamily
  1305.2119}}].

\bibitem{Arnold:2008zu}
P.B.~Arnold and C.~Dogan, \emph{{QCD Splitting/Joining Functions at Finite
  Temperature in the Deep LPM Regime}},
  \href{https://doi.org/10.1103/PhysRevD.78.065008}{\emph{Phys. Rev. D}
  {\bfseries 78} (2008) 065008}
  [\href{https://arxiv.org/abs/0804.3359}{{\ttfamily 0804.3359}}].

\bibitem{Mueller:1999fp}
A.H.~Mueller, \emph{{Toward equilibration in the early stages after a
  high-energy heavy ion collision}},
  \href{https://doi.org/10.1016/S0550-3213(99)00502-7}{\emph{Nucl. Phys. B}
  {\bfseries 572} (2000) 227}
  [\href{https://arxiv.org/abs/hep-ph/9906322}{{\ttfamily hep-ph/9906322}}].

\bibitem{Blaizot:2014jna}
J.-P.~Blaizot, B.~Wu and L.~Yan, \emph{{Quark production,
  Bose\textendash{}Einstein condensates and thermalization of the
  quark\textendash{}gluon plasma}},
  \href{https://doi.org/10.1016/j.nuclphysa.2014.07.041}{\emph{Nucl. Phys. A}
  {\bfseries 930} (2014) 139}
  [\href{https://arxiv.org/abs/1402.5049}{{\ttfamily 1402.5049}}].

\bibitem{APT}
C.~{De Grandi} and A.~{Polkovnikov}, \emph{{Adiabatic Perturbation Theory: From
  Landau-Zener Problem to Quenching Through a Quantum Critical Point}},  in
  \emph{Quantum Quenching, Annealing and Computation, Lecture Notes in Physics,
  Volume 802. ISBN 978-3-642-11469-4. Springer-Verlag Berlin Heidelberg, 2010,
  p. 75}, A.K.K.~{Chandra}, A.~{Das} and B.K.K.~{Chakrabarti}, eds., p.~75
  (2010), \href{https://doi.org/10.1007/978-3-642-11470-0_4}{DOI}
  [\href{https://arxiv.org/abs/0910.2236}{{\ttfamily 0910.2236}}].

\bibitem{Blaizot:2021cdv}
J.-P.~Blaizot and L.~Yan, \emph{{On attractor and fixed points in Bjorken
  flows}},  \href{https://arxiv.org/abs/2106.10508}{{\ttfamily 2106.10508}}.

\bibitem{Bjoraker:2000cf}
J.~Bjoraker and R.~Venugopalan, \emph{{From colored glass condensate to gluon
  plasma: Equilibration in high-energy heavy ion collisions}},
  \href{https://doi.org/10.1103/PhysRevC.63.024609}{\emph{Phys. Rev. C}
  {\bfseries 63} (2001) 024609}
  [\href{https://arxiv.org/abs/hep-ph/0008294}{{\ttfamily hep-ph/0008294}}].

\bibitem{barenblatt1996scaling}
G.~Barenblatt, G.~Barenblatt, B.~Isaakovich, D.~Crighton, M.~Ablowitz, S.~Davis
  et~al., \emph{Scaling, Self-similarity, and Intermediate Asymptotics:
  Dimensional Analysis and Intermediate Asymptotics}, Cambridge Texts in
  Applied Mathematics, Cambridge University Press (1996).

\bibitem{PhysRevLett.64.1361}
N.~Goldenfeld, O.~Martin, Y.~Oono and F.~Liu, \emph{Anomalous dimensions and
  the renormalization group in a nonlinear diffusion process},
  \href{https://doi.org/10.1103/PhysRevLett.64.1361}{\emph{Phys. Rev. Lett.}
  {\bfseries 64} (1990) 1361}.

\bibitem{scaling-QM}
L.~ŠAMAJ, \emph{Evolution of quantum systems with a scaling type
  time-dependent hamiltonians},
  \href{https://doi.org/10.1142/s0217979202013158}{\emph{International Journal
  of Modern Physics B} {\bfseries 16} (2002) 3909–3914}.

\bibitem{Kibble_1976}
T.W.B.~Kibble, \emph{Topology of cosmic domains and strings},
  \href{https://doi.org/10.1088/0305-4470/9/8/029}{\emph{Journal of Physics A:
  Mathematical and General} {\bfseries 9} (1976) 1387}.

\bibitem{Zurek:1985qw}
W.H.~Zurek, \emph{{Cosmological Experiments in Superfluid Helium?}},
  \href{https://doi.org/10.1038/317505a0}{\emph{Nature} {\bfseries 317} (1985)
  505}.

\bibitem{Zurek:1996sj}
W.H.~Zurek, \emph{{Cosmological experiments in condensed matter systems}},
  \href{https://doi.org/10.1016/S0370-1573(96)00009-9}{\emph{Phys. Rept.}
  {\bfseries 276} (1996) 177}
  [\href{https://arxiv.org/abs/cond-mat/9607135}{{\ttfamily
  cond-mat/9607135}}].

\bibitem{KZ2012}
A.~Chandran, A.~Erez, S.S.~Gubser and S.L.~Sondhi, \emph{Kibble-zurek problem:
  Universality and the scaling limit},
  \href{https://doi.org/10.1103/physrevb.86.064304}{\emph{Physical Review B}
  {\bfseries 86} (2012) }.

\bibitem{Mukherjee:2016kyu}
S.~Mukherjee, R.~Venugopalan and Y.~Yin, \emph{{Universal off-equilibrium
  scaling of critical cumulants in the QCD phase diagram}},
  \href{https://doi.org/10.1103/PhysRevLett.117.222301}{\emph{Phys. Rev. Lett.}
  {\bfseries 117} (2016) 222301}
  [\href{https://arxiv.org/abs/1605.09341}{{\ttfamily 1605.09341}}].

\bibitem{Akamatsu:2018vjr}
Y.~Akamatsu, D.~Teaney, F.~Yan and Y.~Yin, \emph{{Transits of the QCD critical
  point}}, \href{https://doi.org/10.1103/PhysRevC.100.044901}{\emph{Phys. Rev.
  C} {\bfseries 100} (2019) 044901}
  [\href{https://arxiv.org/abs/1811.05081}{{\ttfamily 1811.05081}}].

\bibitem{frisch1995turbulence}
U.~Frisch and A.~Kolmogorov, \emph{Turbulence: The Legacy of A. N. Kolmogorov},
  Cambridge University Press (1995).

\bibitem{Hirono:2015rla}
Y.~Hirono, D.~Kharzeev and Y.~Yin, \emph{{Self-similar inverse cascade of
  magnetic helicity driven by the chiral anomaly}},
  \href{https://doi.org/10.1103/PhysRevD.92.125031}{\emph{Phys. Rev. D}
  {\bfseries 92} (2015) 125031}
  [\href{https://arxiv.org/abs/1509.07790}{{\ttfamily 1509.07790}}].

\bibitem{Mace:2019cqo}
M.~Mace, N.~Mueller, S.~Schlichting and S.~Sharma, \emph{{Chiral Instabilities
  and the Onset of Chiral Turbulence in QED Plasmas}},
  \href{https://doi.org/10.1103/PhysRevLett.124.191604}{\emph{Phys. Rev. Lett.}
  {\bfseries 124} (2020) 191604}
  [\href{https://arxiv.org/abs/1910.01654}{{\ttfamily 1910.01654}}].

\end{thebibliography}\endgroup

\end{document}